\pdfoutput=1
\documentclass[sigconf,nonacm]{acmart}
\usepackage{graphicx}
\AtBeginDocument{%
  }

\setcopyright{acmlicensed}
\copyrightyear{2018}
\acmYear{2018}
\acmDOI{XXXXXXX.XXXXXXX}
\acmConference[Conference acronym 'XX]{Make sure to enter the correct
  conference title from your rights confirmation email}{June 03--05,
  2018}{Woodstock, NY}
\acmISBN{978-1-4503-XXXX-X/2018/06}

\usepackage{booktabs}
\usepackage{tabularx}
\usepackage{array}
\usepackage{xspace}
\usepackage{graphicx}
\usepackage{xltabular}
\usepackage{ragged2e}
\usepackage{array}
\usepackage{booktabs}
\usepackage{tabularx}
\newcolumntype{Y}{>{\RaggedRight\arraybackslash}X}

\usepackage[most]{tcolorbox}
\usepackage{listings}
\usepackage{xcolor}

\definecolor{promptbg}{RGB}{248,248,248}
\definecolor{promptframe}{RGB}{190,190,190}

\lstdefinestyle{promptstyle}{
    basicstyle=\ttfamily\footnotesize,
    breaklines=true,
    breakatwhitespace=false,
    columns=fullflexible,
    keepspaces=true,
    showstringspaces=false,
    tabsize=2,
    aboveskip=0pt,
    belowskip=0pt
}

\newtcblisting{promptbox}{
    enhanced,
    breakable,
    listing only,
    listing options={
        style=promptstyle
    },
    colback=promptbg,
    colframe=promptframe,
    boxrule=0.5pt,
    arc=0pt,
    outer arc=0pt,
    left=2mm,
    right=2mm,
    top=2mm,
    bottom=2mm,
    before skip=6pt,
    after skip=8pt
}

\newcommand{\oursystem}{MRPilot\xspace} %
\begin{document}

\title{\oursystem: Supervising and Intervening LLM-Based Multi-Robot Teams through Mixed Reality}

\author{Xiaoran Yang}
\email{xyang49@ncsu.edu}
\affiliation{%
  \institution{North Carolina State University}
  \city{Raleigh}
  \state{North Carolina}
  \country{USA}
}

\author{Xun Qian}
\affiliation{%
  \institution{Google}
  \city{San Jose}
  \state{California}
  \country{USA}
  }
\email{me@xun-qian.com}

\author{Yang Zhan}
\affiliation{%
  \institution{North Carolina State University}
  \city{Raleigh}
  \state{North Carolina}
  \country{USA}
}
\email{yzhan3@ncsu.edu}

\author{Nathan Tran}
\affiliation{%
 \institution{North Carolina State University}
 \city{Raleigh}
 \state{North Carolina}
 \country{USA}}
 \email{nttran4@ncsu.edu}

\author{Ziyi Liu}
\affiliation{%
  \institution{Purdue University}
  \city{West Lafayette}
  \state{Indiana}
  \country{USA}}
  \email{liuziyi1230@gmail.com}

\author{Qiao Jin}
\affiliation{%
  \institution{North Carolina State University}
  \city{Raleigh}
  \state{North Carolina}
  \country{USA}}
\email{qjin4@ncsu.edu}

\renewcommand{\shortauthors}{Yang et al.}

\begin{abstract}
Large language models (LLMs) let users direct heterogeneous multi-robot systems (MRS) through natural language, but make task interpretation, robot assignment, and coordination difficult to inspect and change. Based on a formative study with 12 non-expert users, we developed MRPilot, a mixed reality system organized around four stages of supervision and intervention. MRPilot represents robot-team plans and execution states as structured commitments shared across synchronized situated and overview views. Across four stages, it helps users resolve ambiguous references (Forming), review plans before execution (Reviewing), monitor distributed execution (Following), and make robot-level or team-level changes when problems arise (Repairing). In a within-subjects study with 20 participants in a virtual reality-simulated home, MRPilot reduced workload, increased situational awareness, transparency, trust, and perceived control compared with a conventional LLM-based conversational interface using the same LLM planner and robot capabilities. We provide design implications for multi-scale intervention, adaptive supervision, and calibrated reliance in LLM-based MRS.

\end{abstract}



\begin{teaserfigure}
  \centering
  \includegraphics[width=0.85\textwidth]{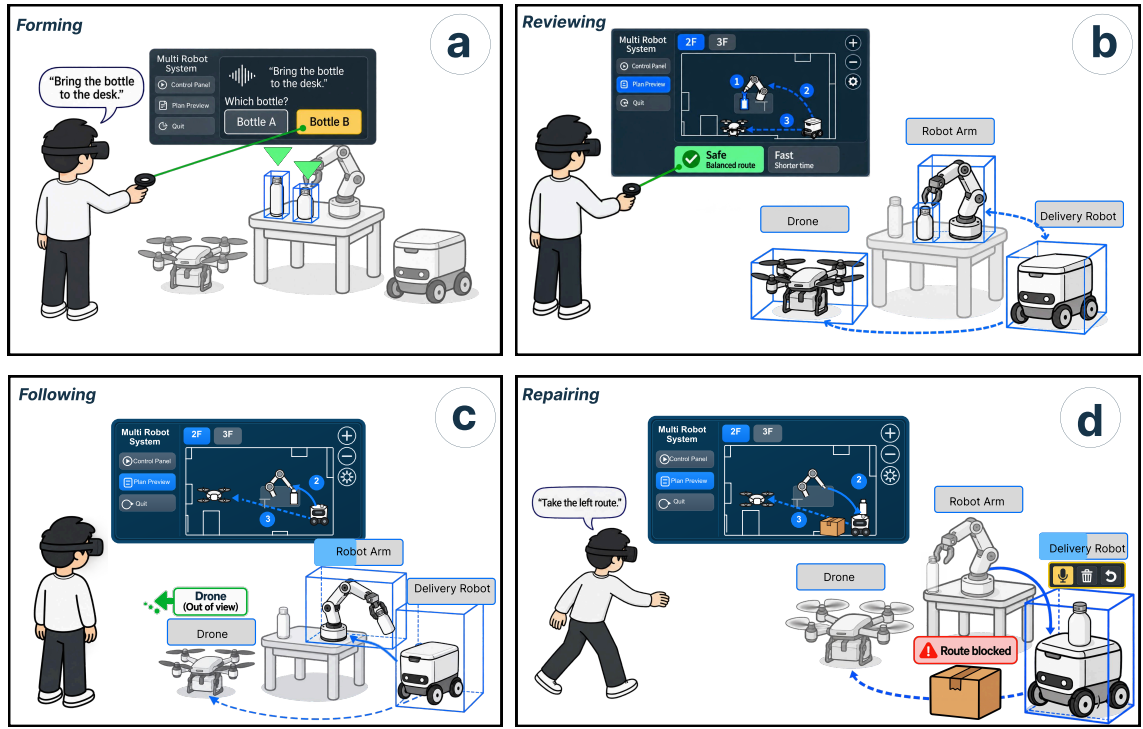}
  \caption{Overview of \oursystem. \oursystem is a MR system for supervising LLM coordinated multi-robot systems (MRS), enabling rapid intervention in MRS tasks through spatial interaction. In a task execution workflow, the user first issues an instruction in natural language. (a) \oursystem detects ambiguities in the instruction, highlights non-unique elements in the environment, and displays them in the interaction panel. (b) \oursystem visualizes the current candidate plans through the MR view and Mini-Map for user selection. (c) The MR view and Mini-Map continuously update robot positions and plan information during execution. (d) When unexpected events requiring intervention occur during execution, the user can quickly inspect relevant information and intervene through the MR view.}
  \label{fig:teaser}
\end{teaserfigure}

\maketitle

\section{Introduction}
Multi-robot systems (MRS) consist of multiple robots that coordinate to complete shared or related tasks~\cite{rizk2019cooperative, parker2008handbook}. 
Traditional MRS interfaces often require users to specify tasks, assign robots, define routes or waypoints, and manage execution through system-specific controls~\cite{chen2010supervisory, chen2014human}. Large language models (LLMs) allow users to instead describe high-level goals in natural language. The system can then interpret the instruction, decompose it into subtasks, assign robots based on their capabilities, and produce a coordinated execution plan~\cite{kannan2024smart, liu2025coherent, mandi2024roco, chen2024scalable}. 
However, transferring this coordination work to an LLM can make task interpretation, robot allocation, and planning decisions less visible to users~\cite{kim2024understanding, ren2023robots}. 
Because LLMs may misinterpret underspecified instructions or generate actions that do not fit the current context~\cite{lee2025veriplan, kim2024understanding}, human supervision remains necessary for supplying context, evaluating plans, and correcting execution~\cite{chen2014human, parasuraman2000model}. %

Effective human supervision and intervention in LLM-based MRS remain challenging. First, LLM-generated plans may appear plausible while containing errors, such as assigning tasks to robots without the required capabilities or omitting necessary preconditions~\cite{huang2022inner, kannan2024smart}. Users therefore need to inspect and revise task interpretations, robot assignments, action dependencies, and planning rationales before and during execution. Second, multi-robot execution is spatially distributed and interdependent. Robots may operate outside the user’s field of view~\cite{chen2007human, roldan2017multi, walker2023virtual}, and understanding one robot's behavior often requires relating it to the broader execution structure~\cite{korsah2013comprehensive, nunes2017taxonomy}. In heterogeneous MRS, this also requires users to relate each robot’s capabilities to its assigned role~\cite{rizk2019cooperative, zhang2025lamma}. Such plans combine spatial information about robots, objects, and locations; temporal relationships among actions; and contextual constraints~\cite{korsah2013comprehensive, gombolay2018fast}. These relationships can be difficult to inspect and modify through chronological text-based interfaces~\cite{ge2025gencomui, yang2026understanding}.%

Mixed Reality (MR) may support the supervision of spatially distributed MRS by embedding task states, planned actions, and robot-specific information in the physical environment, directly linking them to the entities and locations they describe~\cite{suzuki2022augmented, walker2023virtual, walker2018communicating}. Such situated representations can reduce the need to mentally map abstract interface information onto the environment~\cite{hedayati2018improving, szafir2021connecting}. However, spatial embedding can also increase visual search and perceptual costs, whereas 2D overview representations can better support comparison and global decision-making~\cite{liu2026can}. %
Existing interfaces provide limited support for connecting the LLM’s interpretation and planning decisions with the spatial and temporal structure of multi-robot execution. It remains unclear how an interface can support users in inspecting instruction interpretation, evaluating coordination plans, maintaining awareness during execution, and making targeted corrections. 

To identify the specific breakdowns and user needs that such an interface should address, we first conducted a formative study with 12 participants interacting with an LLM-based MRS. Based on the findings, together with prior work on situated visualization and overview visualization, we developed \textbf{\oursystem, an MR system for spatially situated supervision and intervention of LLM-based MRS}. \oursystem represents planning decisions as structured commitments that connect each robot action to its referents, location, dependencies, and rationale across the LLM planner, interface, and robot execution. Its workflow supports users in resolving ambiguous references during Forming, inspecting and modifying plans during Reviewing, monitoring spatially distributed execution during Following, and making robot- or team-level corrections during Repairing. 
We compared \oursystem with a conventional text-and-voice LLM-based interface in a within-subjects study with 20 participants in a Virtual Reality (VR)-simulated home. The results showed that \oursystem reduced task completion time compared with the conventional interface. It also showed improvements in transparency, trust, situation awareness, and workload. The qualitative results further showed that intuitive interaction, spatial visualization, and visualization of the execution process were three key factors through which \oursystem improved users’ situation awareness and trust, while also introducing corresponding challenges. At the same time, these mechanisms introduced some interaction costs, field-of-view limitations, and issues with explainability.

This work makes three contributions:
\begin{itemize}
    \item Empirical findings from a formative study characterizing how non-expert users supervise LLM-based heterogeneous MRS and the difficulties they encounter during this process.
    \item \oursystem, an MR system that supports spatially situated supervision and intervention in LLM-based MRS through coordinated situated and overview representations.
    \item Findings from a controlled within-subjects study comparing MRPilot with a conversational interface for supervising and intervening in an LLM-based heterogeneous MRS.
\end{itemize}

\section{Related Work}
\subsection{LLM-based Heterogeneous Multi-Robot System}
Heterogeneous multi-robot systems (MRS) combine complementary morphologies and capabilities and are suited to complex real-world tasks~\cite{tranzatto2022cerberus}.
However, heterogeneity complicates interaction. Allocation must account for each robot's capabilities rather than workload alone~\cite{wang2008assessing}, and plans must be coordinated across robots with different action spaces~\cite{wang2007human}. Users must therefore integrate information from multiple robots, increasing workload and reducing situation awareness~\cite{roldan2017multi, dahiya2023survey}. Controllers and task specifications also typically require expert programming for each task, which is slow and hard to adapt during deployment~\cite{strobel2024llm2swarm}.
LLMs reduce this burden but introduce new problems. First, users act only as one-time instruction providers at the start of a task~\cite{li2026large}, and the capability-matching process is invisible to them~\cite{wang2026explainable}. Second, LLM errors are hard to correct in time~\cite{feng2026cocoa, kim2024understanding}: models may produce hallucinated or inconsistent plans, refer to nonexistent objects, or propose infeasible actions~\cite{rivera2025conceptagent, yao2026bridging} that look plausible until inspected~\cite{ma2026maestro}. Heterogeneity adds cross-robot state errors, such as one robot pursuing a subgoal another has already completed~\cite{yu2026dynahmrc, liu2025coherent}; once execution begins, such errors accumulate with few chances to intervene. Third, whereas traditional interfaces expose robot states through camera feeds and maps~\cite{dahiya2023survey}, LLM-based systems drive behavior through model decisions whose reasoning is hidden~\cite{das2021explainable}, limiting failure anticipation and diagnosis and reducing trust~\cite{yildiz2026show}. 
Existing feedback mechanisms allow robots to report failures to the LLM, which can reassign tasks or request operator help~\cite{li2025hmcf, borate2026comuros, hasegawa2026reducing}. These mechanisms primarily support system-initiated recovery, involving users only after the system detects a failure~\cite{hasegawa2026reducing}. To support continuous human-in-the-loop monitoring and intervention in LLM-based heterogeneous MRS, we begin with user needs and focus on small-scale robot teams of 2--5 robots with 2--5 morphologies, typical of current systems~\cite{kannan2024smart, mandi2024roco}. %

\subsection{Human Supervision and Intervention in Multi-Robot Systems}

Human oversight of MRS involves two related activities: supervision and intervention~\cite{sheridan1992telerobotics}. Supervision requires operators to acquire and interpret current system states and anticipate future ones~\cite{parasuraman2000model}. Traditional MRS interfaces support this process by presenting each robot's video and status~\cite{wang2007human}, while exposing team plans and decision rationales can improve supervisory performance and support calibrated trust without increasing workload~\cite{mercado2016intelligent, chen2014human}. LLM-based MRS add another layer because plans are generated from natural-language instructions during interaction rather than specified in advance~\cite{kannan2024smart, mandi2024roco}. Operators must therefore understand how the model interprets instructions, assigns robots, and coordinates their actions. Some systems expose parts of this process through inter-robot dialogue~\cite{mandi2024roco}, but these logs grow with team size~\cite{chen2024scalable} and lack structured views~\cite{epperson2025interactive}. 
Intervention refers to corrective actions humans take to modify or redirect autonomous behavior~\cite{steinfeld2006common}. It includes teleoperating a robot~\cite{sheridan1992telerobotics}, taking over a subtask~\cite{sellner2006coordinated, scerri2002towards}, reallocating tasks, or invoking predefined plays~\cite{miller2007designing, parasuraman2005flexible}, and may be initiated by the operator or system~\cite{sellner2006coordinated, li2025hmcf}. LLM-based systems support task-level intervention by allowing users to confirm or modify assignments~\cite{li2025hmcf} and provide corrective natural-language feedback~\cite{ji2026genswarm}. However, intervention remains concentrated before execution or after failure~\cite{li2025hmcf, ji2026genswarm}, with limited support for targeted corrections during execution. 
\oursystem addresses this gap by presenting team plans and execution states through linked situated MR and overview views while supporting robot- and team-level intervention through spatial selection, direct controls, and natural-language commands.

\subsection{Mixed Reality in Multi-Robot System}

Mixed reality (MR) registers virtual content with the physical world~\cite{azuma1997survey, cao2019ghostar}, allowing robot states and planned actions to be presented at their referenced locations~\cite{walker2018communicating, yanco2004beyond}. Spatial cues such as trajectories, goal markers, and anchored labels can communicate robot intent and support situation awareness~\cite{suzuki2022augmented, walker2018communicating, hedayati2018improving, debernardis2013text}. MR also supports situated intervention by allowing operators to refer to robots, objects, and locations through pointing or gaze and to create or edit waypoints in place~\cite{williams2019deictic, suzuki2022augmented}.
However, MR provides an egocentric and partial view, whereas MRS supervision requires a persistent overview of the robot team~\cite{rule2012designing}. Spatial overlays can also become cluttered as robot numbers increase~\cite{rule2012designing}, head-mounted displays are poorly suited to dense text~\cite{rzayev2018reading}, and map views can better communicate global task structure~\cite{liu2026can}. 
This combination is particularly relevant to LLM-based MRS. Their textual plans are repeatedly revised~\cite{mandi2024roco} and accumulate during execution~\cite{chen2024scalable}, allowing the current plan to become buried in a linear text stream~\cite{jiang2023graphologue}. Because these plans encode spatial information~\cite{kannan2024smart, mandi2024roco}, MR can connect plan steps to their objects and locations, helping users identify inconsistencies between the plan and environment~\cite{chandan2021arroch, wang2023explainable}. MR can also ground open-ended instructions through speech and spatial reference~\cite{williams2019deictic} and bind dialogue-based corrections~\cite{kim2024understanding} to selected robots or plan elements~\cite{krupke2018comparison}. Yet prior MR robotics work focuses mainly on individual robots~\cite{suzuki2022augmented, walker2018communicating, hedayati2018improving} or MRS with predefined behaviors~\cite{adekoya2026horus}. LLM-based MRS instead remain largely text-based~\cite{li2026large}, requiring users to infer team states and robot identities from dialogue~\cite{mandi2024roco, li2025hmcf} without direct spatial links to the robots or environment~\cite{yang2026mave}.
Building on these complementary strengths, we investigate how linked overview and situated MR representations can support plan inspection, execution monitoring, and targeted intervention in LLM-based MRS.

\section{Formative Study}

To inform our system design, we conducted a formative study guided by the following research question: \textbf{What challenges do users encounter when directing an LLM-based heterogeneous MRS through natural-language instructions in spatial tasks?} We focused on spatial tasks involving heterogeneous robots, where geometry, limited visibility, and interdependent actions complicate coordination. 

\subsection{Participants, Procedure, Data collection and analysis}
We recruited 12 participants (ages 22–32, $M = 25.25$, $SD = 2.89$) via a university mailing list, compensated with gift cards. They varied in familiarity with AI tools (2 high, 4 moderate, 6 low) and MRS (3 none, 3 basic, 6 relatively high). The study was approved by our institutional review board. The platform was a VR multi-robot simulation testbed built in Unreal Engine 5 and run on an Oculus Quest 3S. A virtual house contained three heterogeneous robots: a 6-degree-of-freedom arm for manipulation, a quadruped for ground transport, and a quadrotor drone for aerial transport. Following a standard MRS workflow~\cite{rizk2019cooperative}, an LLM-based planner parsed natural language instructions, identified target objects, decomposed them into subtasks, and assigned these dynamically based on robot states and perception. The interface displayed robot types, generated subtasks, and real-time execution status.

Each study session lasted about 70 minutes. After consent, a demographic questionnaire, and a brief VR training on the robots and voice interaction, participants completed two tasks in counterbalanced order: retrieving food from outside the house (Task 1) and placing a book near a second-floor bookshelf (Task 2). Both were adapted from manipulation and retrieval benchmarks in prior LLM-based MRS work~\cite{ahn2022can, huang2022inner, mandi2024roco, wu2023tidybot,zhang2024building}. Since spatial visibility affects task behavior~\cite{lapeyre2011using, patel2021transparency}, the robots' execution was partially hidden in Task 1 and fully visible in Task 2. Participants could phrase instructions freely. The researcher observed and recorded reactions, offering operational help but no strategic guidance. Each session ended with a 30-minute semi-structured interview on participants' experience, prompting strategies, and difficulties. Transcripts were analyzed using thematic analysis~\cite{ahmed2025using, muller2014curiosity}: the first author developed an initial coding scheme, and all authors then collaboratively refined the codes into the final themes reported below.

\subsection{Findings}

\subsubsection{\textbf{F1: Users are often unaware that their prompts are underspecified, because the system gives no feedback before acting.}}

Participants began with high-level goal descriptions and assumed the system understood the task as they did. Because the system never surfaced ambiguities or requested clarification, it proceeded directly to planning and execution, so problems appeared only when the outcome conflicted with expectations: \textit{``ambiguous prompts could sometimes lead to results that deviated from the original intent''} (P9). Participants therefore diagnosed retrospectively and compensated through repeated revision---adding detail, assigning subtasks to specific robots (P5, P11), or restructuring instructions to \textit{``describe things step by step''} (P3)---typically when \textit{``the result was very slow, or the result was incorrect''} (P5), or the goal \textit{``was achieved but was clearly inefficient''}(P7). Opacity compounded the uncertainty: \textit{``Revising the initial prompt is most valuable when I cannot observe or understand the complete execution process of the robot system''} (P3). Participants instead wanted reasoning or feedback before execution, such as \textit{``a feedback mechanism to confirm whether what I wanted it to do had been completed''} (P2) or for the system to \textit{``show its thought process''} (P8). In most cases, they recognized a flawed prompt only after paying the cost of a failed or inefficient execution.

\subsubsection{\textbf{F2: Direct commitment to an unpreviewed plan makes robot coordination difficult to assess.}}
When generating a plan, the system typically committed to a single plan and proceeded directly to execution. As a result, participants often discovered only during or after execution that the task allocation did not match their expectations. The selected plan also frequently differed from what participants considered a better alternative. For example, P9 believed the drone could complete the task alone, while P11 observed,  \textit{``Something that could clearly be done by one robot was unnecessarily split across three robots, which made the whole system feel very rigid.''} 
Modifying a plan required rewriting the prompt and restarting the entire task, which participants found cumbersome. They wanted the system to present the plan explicitly before execution, for example by restating the actions it was about to perform for confirmation (P2) and showing the rationale behind its planning decisions (P8). 
Their assessments of plans considered whether the number of robots matched the task (P10) and whether assignments reflected robot capabilities (P4). These responses suggest that participants wanted to review candidate plans and select among alternatives before committing it to execution.

\subsubsection{\textbf{F3: Distributed multi-robot execution creates a gap in spatial awareness that textual information does not fully address.}}

With three heterogeneous robots operating in parallel across indoor and outdoor spaces, participants found it difficult to track which robot was doing what and where: \textit{``I wasn't really sure where exactly they handed things off to each other''} (P12). Text-based status displays did not resolve this---\textit{``all I got were text prompts, which did not help me understand the current execution progress intuitively''} (P4)---and competed with the spatial task for attention budget: \textit{``There were too many elements, and the information density was very high''} (P3); \textit{``The panel blocked my view of the robots, and I had to keep looking back at it, which made it difficult to understand how the entire robot system was operating''} (P12). Rather than concentrating information in a panel, participants wanted it placed back into the spatial environment: trajectories shown \textit{``directly in the scene instead of limiting them to a small panel''}(P4), text anchored \textit{``directly to the corresponding robot''} (P8), \textit{``a top-down view on the side so I can see how each robot is moving''} (P8), and \textit{``the real-time view of each robot''} (P3). Awareness of distributed execution thus needs the spatial context of the robots themselves; text alone was insufficient and at times distracting.

\subsubsection{\textbf{F4: High-level prompting is a poor fit for localized intervention.}}

Participants' corrective intentions were usually local---changing the task assigned to one robot, adjusting the execution order of a subtask, or correcting a single step. Yet natural language, the only control channel, forced them to reformulate and reissue the entire instruction (P3), risking replanning of a task that was mostly appropriate.
Participants worked around this limitation by embedding component-specific instructions within a global prompt. For example, they directly referred to a particular robot: \textit{``Can the drone bring some food back from outside?''} (P11). Others responded by adding increasing amounts of detail to their initial instructions (P5), effectively micromanaging the planner through natural language.
These workarounds further reveal a mismatch in interaction granularity. Participants’ corrective intentions were often local and targeted, while the existing interaction mechanism forced them to operate at the level of the entire task. As a result, local intervention became both effortful and difficult to perform precisely.

\subsection{Design Goals}

We synthesized these findings into three design goals for a system that better supports intervention and decision making in spatial tasks with LLM-based MRS.

\textbf{DG1: Make task interpretation and coordination decisions inspectable before execution.}
Participants often discovered interpretation errors only after execution and had limited opportunities to assess robot allocation beforehand (F1, F2), consistent with prior work showing that LLM-generated plans can appear plausible while containing hidden errors~\cite{zhao2024explainability, valmeekam2023planning}. The system should expose how it resolves objects, locations, constraints, and user intent, together with the resulting robot assignments, actions, dependencies, and rationales. Users should be able to inspect and revise these decisions before confirming the plan.

\textbf{DG2: Maintain coordination continuity across global and situated contexts.}
Participants wanted to assess task allocation before execution and track robot roles, locations, and handoffs during execution, including activity outside their field of view (F2, F3). Prior work shows that situated representations can connect information to physical entities, while overview representations can support awareness of distributed robot activity~\cite{suzuki2022augmented, walker2023virtual, walker2018communicating}. The system should therefore preserve consistent relationships among robots, actions, referents, paths, dependencies, and execution states across planning, execution, and linked global and situated views.

\textbf{DG3: Support context-bound intervention with visible plan consequences.}
Participants often wanted to correct a specific assignment, action, object, or location, but the existing interface required broader task-level instructions and replanning (F4). Prior work shows that robot actions remain connected through dependencies and handoffs, so changing one component may affect the wider coordination plan~\cite{korsah2013comprehensive, nunes2017taxonomy, chen2014human}. The system should support interventions scoped to selected task elements, preview their effects on related assignments, paths, dependencies, and action order, and preserve unaffected parts of the plan.

\section{\oursystem}
We present \oursystem, an MR system for supervising and intervening in LLM-coordinated MRS. \oursystem supports a four-stage workflow in which users resolve ambiguous references, review candidate plans, monitor distributed execution, and make robot- or team-level corrections.
To maintain consistency across this workflow, \oursystem represents information exchanged among the user, LLM planner, interface, and robot execution as structured \emph{commitments}. A UI translation layer renders these commitments in the linked Mini-Map and MR views, while an action translation layer maps them to positions and paths for robot execution.

\subsection{\oursystem Framework}
\subsubsection{System Workflow}\label{sec: wfw}

To translate our design goals into an interaction workflow, \oursystem organizes supervision and intervention into four stages: \textbf{Forming}, \textbf{Reviewing}, \textbf{Following}, and \textbf{Repairing}. In \textbf{Forming}, the system detects ambiguous elements in user instructions and externalizes them as visual spatial markers, addressing DG1. In \textbf{Reviewing}, the system presents relationships among plans, including dependencies, conflicts, and coverage, in the user interface for selection, addressing DG1 and DG2. \textbf{Following} addresses DG2 by visualizing the execution process and robot states, allowing problems during execution to be shown in real time for users to inspect. \textbf{Repairing} also addresses DG2 and DG3. When intervention is needed, the system localizes the problem to a specific robot, allows users to make local modifications through the interface, and lets them preview how the repair affects the confirmed plan. Fig.~\ref{fig:teaser} illustrates an example interaction across these four stages.

\subsubsection{Commitment-Based System Architecture}\label{sec:sst}

\paragraph{\textbf{Data Representation}}
Two technical challenges arise in supporting interaction between users and the system. First, the data formats are different. Users provide natural language input, robot teams receive spatial position information, and the LLM processor receives structured input. These inconsistent data formats can reduce interaction efficiency. Second, our system workflow consists of continuous interaction, so information transfer needs to remain consistent throughout the workflow. Inconsistent information representations can also lead to efficiency problems. To address these challenges, we designed a unified data structure, \emph{commitments}. Data exchanged among users, robot teams, and the LLM is now represented through commitments. Our workflow also uses \emph{commitments} as the basic data unit for implementing its functions. It is a structured array of eight elements. 
We use the task “the delivery robot carries the red cup out of the kitchen and hands it to the robot arm” as an example to illustrate how a natural-language instruction is progressively constructed into our \emph{commitments}.

\begin{itemize}
  \item \emph{id}: a unique identifier for the commitment. It lets other commitments (via \emph{waits}) and user interventions refer to a specific commitment unambiguously (DG3). Example: \texttt{c2}.
  \item \emph{robot}: the robot that performs the task. It makes task allocation explicit so users can check who does what before execution (DG1). Example: \texttt{Robot Arm}.
  \item \emph{action}: the action the robot executes, drawn from its supported set (\textsc{PICK UP}, \textsc{DROP OFF}, \textsc{TRANSPORT}). Restricting the vocabulary to executable actions keeps every commitment both interpretable and directly executable (DG1). Example: \textsc{TRANSPORT}.
  \item \emph{referent}: the object involved in the action. It exposes how the system grounded the user's reference (e.g., ``the cup'') to a concrete object, so misinterpretations can be caught early (DG1). Example: \texttt{red cup}.
  \item \emph{place}: the destination of the action. It anchors the commitment in the situated environment, linking the global plan to the location where execution occurs (DG2). Example: \texttt{living room}.
  \item \emph{receiver}: the robot that receives the object when the action is a handoff; empty for the final task. It makes inter-robot dependencies explicit so coordination is preserved when a plan spans multiple robots (DG2). Example: \texttt{Delivery Robot}.
  \item \emph{waits}: the preceding commitments that must complete before this one begins; empty when there is no prerequisite. It encodes plan ordering, so that modifying one commitment reveals which downstream commitments are affected (DG3). Example: \texttt{[c1]}.
  \item \emph{rationale}: the reason the commitment was generated. It lets users judge whether the interpretation and plan are correct before execution and understand the consequences of changing them (DG1, DG3). Example: ``Robot Arm cannot enter the kitchen, so Delivery Robot carries the cup to the hallway and hands it to Robot Arm.'' And actions that cannot be directly represented by the elements in \emph{action}, such as slowing down.
\end{itemize}

After integrating the elements above, the final \emph{commitment} is as follows:
\begin{quote}
\texttt{[c2, Delivery Robot, TRANSPORT, red cup, living room, Robot Arm, c1, ``Robot Arm cannot enter the kitchen, so Delivery Robot carries the cup to the living room and hands it to Robot Arm.'']}
\end{quote}

\paragraph{\textbf{Core Modules}} Commitments serve as the shared data unit across three core modules: the LLM Processor, UI Translation, and Action Translation. The LLM Processor creates and updates commitments during task planning and replanning. UI Translation renders them as interface elements for users, while Action Translation converts them into spatial positions and paths for robot execution.

\begin{itemize}
    \item \textbf{LLM Processor}: consists of the LLM planner, scene information, and prompts describing robot capabilities. When a user issues a task, the planner receives the user's \emph{commitments} and generates an initial plan represented as a set. Each commitment specifies the task assignment, action order, and dependencies for the corresponding robot. Representing the plan as \emph{commitments} allows users to inspect task interpretation and coordination before execution (DG1).
    \item \textbf{UI Translation}: converts commitment sets into visual elements in the interface. It uses the \emph{receiver}, \emph{place}, and \emph{robot} fields in the commitment sets to identify dependencies and draw lines or bounding box in the user interface. UI translation is used when commitment sets are received from the backend and when the plan is modified. It allows users to inspect the plan before execution (DG1), view the same plan in both the global Mini-Map and the situated MR view (DG2), and preview the effects of modifications (DG3).
    \item \textbf{Action Translation}: converts the structured instructions in each commitment into position information that robots can execute. It first parses the \emph{place}, \emph{robot}, and \emph{receiver} fields and identifies the positions of the corresponding entities from the frontend labels. For example, for \emph{the robot arm pick up the bottle}, the method matches \emph{robot arm} and \emph{bottle} with the frontend labels and stores their positions. Other objects involved in the plan are bound in the same way. After all positions are obtained, the method computes the path, start point, and end point, then assigns the corresponding path information to each robot through the \emph{robot} field. By continuously binding \emph{commitments} to the robots' actual positions and paths, Action translation maintains consistency between the global plan and the current spatial context (DG2). Changes to path preference locations are also directly reflected in the computed paths (DG3).
\end{itemize}

\begin{figure}
    \centering 
    \includegraphics[width=1\columnwidth]{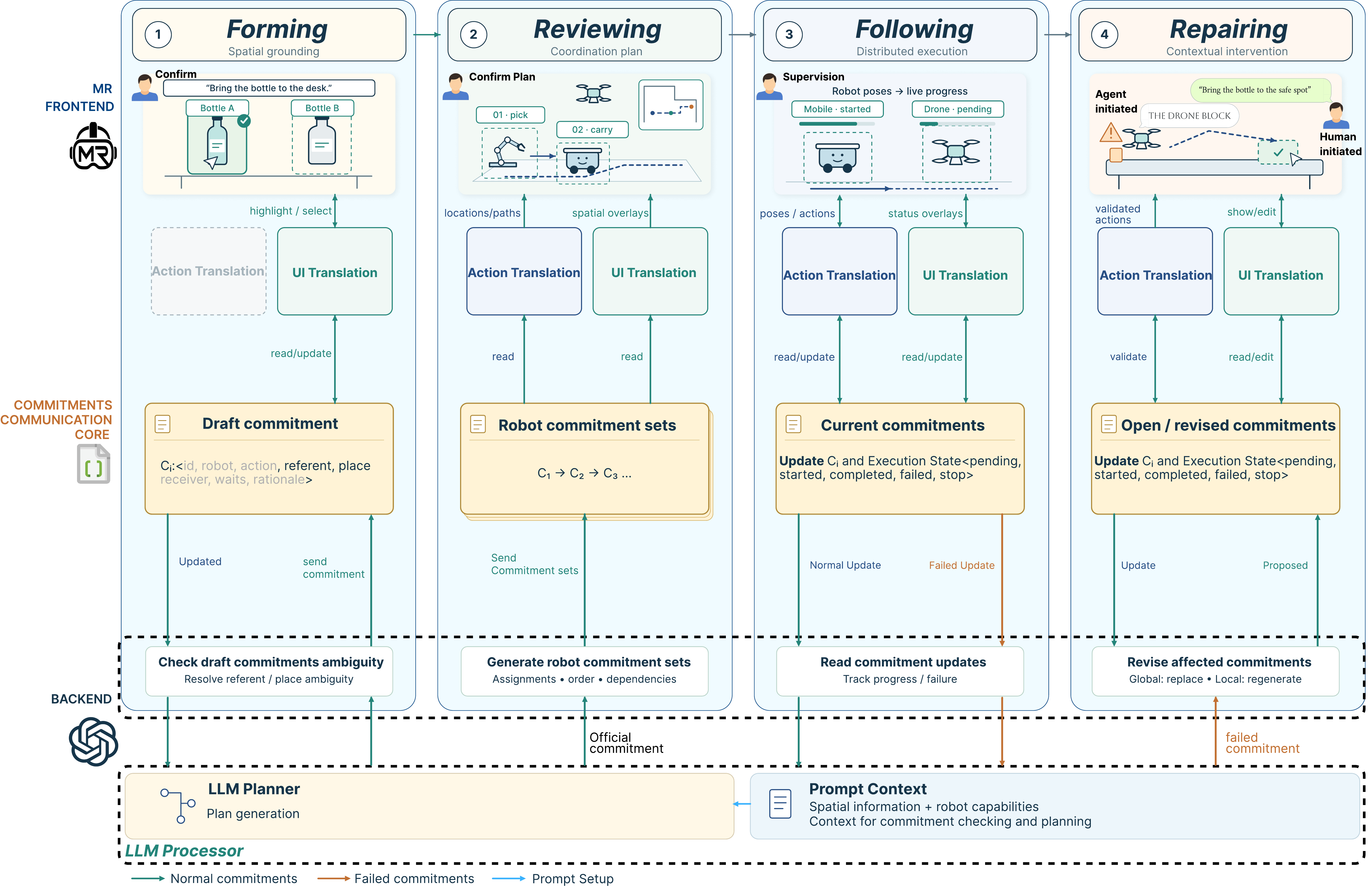}
    \caption{\oursystem system Front-end and back-end communication structure. Frontend consists of the user interface, UI translation, and action translation, which translate the \emph{commitments} generated by the backend into information that can be understood by both the robot team and the user. The backend consists of an LLM processor, which includes an LLM planner together with prompts describing spatial information and robot capabilities. Communication between the frontend and backend is based on structured commitments as the basic data unit, which are transmitted in JSON files.}
    \label{fig:SS1}
\end{figure}

\subsubsection{Commitment Flow and State Transitions}
For each step of the workflow, we implemented corresponding functions through communication between the frontend and backend (Fig.~\ref{fig:SS1}). In \textbf{Forming} (Fig.~\ref{fig:SS1}-1), the user enters a natural language instruction such as "Bring the bottle to the desk." \oursystem extracts the \emph{referent} (\emph{bottle}) and \emph{place} (\emph{desk}), stores them in a draft commitment, and sends it to the LLM Processor. If an element is ambiguous—e.g., both Bottle A and Bottle B exist—the LLM Processor returns the commitment and UI translation displays both candidates for the user to choose from. The updated draft is resubmitted, and once it passes the check it becomes a formal commitment used to generate candidate plans. In \textbf{Reviewing} (Fig.~\ref{fig:SS1}-2), the LLM planner generates two candidate plans from the formal commitment. Each plan contains three \emph{commitments}, one each for the drone, robot arm, and delivery robot. The commitment sets of Plan A and Plan B are sent to the frontend, where action translation converts them into position information for robot execution and UI translation visualizes them for user preview. In \textbf{Following} (Fig.~\ref{fig:SS1}-3), the execution state changes to \emph{started} once execution begins. Action translation continuously records the current position and updates the commitment. When the \emph{place} value matches the predefined path endpoint, the state is set to \emph{completed}; if the update times out, the state is set to \emph{failed} and the current \emph{commitments} are sent to the backend to start intervention. In \textbf{Repairing} (Fig.~\ref{fig:SS1}-4), intervention can be initiated by the system or the user. In system-initiated intervention, e.g., when a drone is blocked by an object, the timer expires and the state becomes \emph{failed}; the \emph{commitments} are sent to the LLM Processor, the affected commitment becomes editable, and UI translation updates the display. In user-initiated intervention, the user identifies an unsuitable assignment, path, action, or result—e.g., deciding the delivery robot should take a shorter route—and sets the execution state to \emph{stop}; the modified \emph{commitments} are then sent to the LLM Processor to generate a new plan. Intervention can also be global or local: global intervention has the LLM Processor generate a new commitment set, while local intervention directly modifies specific elements of the affected robot's commitment. When the drone is blocked, for example, global intervention may coordinate the delivery robot and the drone to clear the obstruction, whereas local intervention only changes the \emph{place} field of the drone's commitment.

\subsection{User Interface Design}
\label{sec:interaction-design}

\oursystem presents and updates \emph{commitments} through two linked views: a team-level \emph{Mini-Map} and a situated \emph{MR view}. Both views are generated from the same \emph{commitments} and remain synchronized throughout the workflow. The Mini-Map provides a global view of task allocation and coordination across the robot team, while the MR view shows where and how these \emph{commitments} are carried out in the environment. We next describe how this interface is designed across the four stages of the workflow and the functions supported at each stage.

\subsubsection{Forming: Resolving Spatial Ambiguity}
\oursystem highlights ambiguous elements in the draft commitment in the spatial environment, as shown in Fig.~\ref{fig:SD1-1}:1-A, 1-B. The MR view displays all current candidates using bounding boxes, while the Mini-Map highlights the marker of the corresponding robot to indicate an active interaction task. The candidate items are also shown as selectable controls in the instruction panel. When the user hovers over an object's button, the corresponding object is highlighted.

\subsubsection{Reviewing: Previewing and Selecting Plans}
\oursystem presents the generated coordination plans as previews in both views. The Mini-Map uses a top-down view to show robot dependencies and planned paths, with numbers and the colors of robot markers indicating the execution order(Fig.~\ref{fig:SD1-1}: 2-A). The MR view shows the robot paths from a first-person perspective, together with a robot label and bounding box for each robot(Fig.~\ref{fig:SD1-1}: 2-B, 2-C). Users can preview the plan in either view and choose one according to their preference (Fig.~\ref{fig:SD1-1}: 2-D).

\subsubsection{Following: Monitoring Distributed Execution}
\begin{figure}
    \centering 
    \includegraphics[width=0.8\columnwidth]{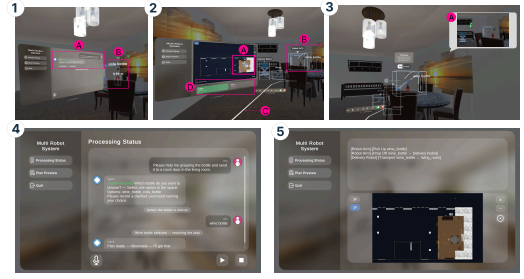}
    \caption{1: Forming stage UI. 1-A: Users select an ambiguous object through the chat panel. 1-B: In the MR view, spatially ambiguous objects are highlighted with bounding boxes, labeled with their names and relative distances. 2: Reviewing stage UI. 2-A: Plan preview in the Mini Map. Each robot has a marker, colored numbers 1 and 2 indicate the execution order, and the stationary robot arm is shown in white. 2-B: During plan preview, the MR view shows each robot with a bounding box and label. 2-C: White lines show the robot team's predicted execution paths.3: Following stage MR view. Each robot's label shows its execution progress, the spatial card shows the current task content, and bounding boxes highlight all robots and objects involved in the task. Execution paths update in real time with robot positions. 3-A: When an event requiring user interaction falls outside the user's field of view, a green guidance arrow at the edge of the view points toward its source.4:Baseline-Processing Status interface, which is similar to the ChatGPT chat interface. Users control the robot team through voice input using the recording button. 5: Baseline- Plan Preview interface. The upper section uses text to show dependencies among multiple robots and task details. The lower section shows a layout map with the room layout and object names.}
    \label{fig:SD1-1}
\end{figure}

\oursystem updates robot paths in real time on both the Mini-Map and MR view. In the MR view, the robot label uses a progress bar to show the current execution progress, while the Spatial Status Card displays the robot's current action(Fig.~\ref{fig:SD1-1}:3). For example, when the Robot Arm is grasping an object, it displays \emph{PICK UP}. For robots outside the user's field of view, the MR view displays green guidance arrows at the edge of the view to draw attention to them(Fig.~\ref{fig:SD1-1}:3-A). The Mini-Map dynamically shows the markers and paths of all robots, together with the dependency structure and execution order of the plan, as shown in Fig.~\ref{fig:SD1-2}: 1.

\subsubsection{Repairing: Revising Plans During Execution}
\begin{figure}
    \centering 
    \includegraphics[width=0.8\columnwidth]{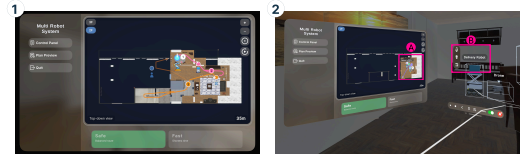}
    \caption{1 Mini Map during task execution, with markers and paths updated in real time. 2-A: In the Mini Map, tapping a robot marker opens its action bar (2-A). 2-B: In the MR view, clicking the robot label opens the same action bar. The action bar supports per-robot adjustments: replanning, removing the robot from the plan, and reverting it to its previous position.}
    \label{fig:SD1-2}
\end{figure}

\oursystem allows users to modify the plan through either the Mini-Map or MR view. In the Mini-Map, clicking a robot marker opens the corresponding Action Bar in the MR view, as shown in Fig.~\ref{fig:SD1-2}: 2-A, Fig.~\ref{fig:SD1-2}:2-B. The record button on the Action Bar supports local intervention for the selected robot. The delete and revert buttons remove the robot from the current plan or reset it to the previous checkpoint. Users can also use the record button on the Control Panel for global intervention. After a modification is made, the paths and dependencies shown in both the MR view and Mini-Map are updated in real time. Users select \emph{Play} on the Toolbar to accept the modification, while \emph{Stop} pauses execution and allows a new intervention.

\subsection{System Implementation}
\oursystem was implemented as an MR visualization and interaction system for spatially situated multi-robot supervision. In the controlled study, we deployed both the interface and the robots in a VR simulation environment, allowing participants to interact with the system from a first-person immersive perspective while preserving the spatial relationships on which the interface design relies. The frontend was developed in Unity 6, while the backend was implemented in TypeScript and deployed as a serverless Cloudflare Worker\footnote{developers.cloudflare.com/workers}. Information between the frontend and backend was exchanged through structured JSON files. The application uses OpenXR together with Meta's XR SDK and Interaction SDK\footnote{developers.meta.com/horizon/documentation/unity} to support immersive viewing, locomotion, and ray-based selection of robots, scene objects, and interface elements. Robot task execution is simulated on the Unity client. \emph{commitments} generated by the backend are mapped to scene objects, target locations, and robot controllers, and are then distributed to the corresponding simulated robots. The robotic arm performs pick-and-place actions through inverse kinematics implemented with the Flange package\footnote{github.com/Preliy/Flange}, while the mobile robots and drone use Unity's NavMesh Agent to simulate movement in the environment.

The backend LLM Processor was built using the GPT-5.4 API with the temperature set to 0. Users' voice instructions are first transcribed into text using gpt-4o-mini-transcribe\footnote{platform.openai.com/docs}. The LLM planner then generates commitment sets for the multi-robot task plan based on a declarative model of the current environment. This environment model includes the currently available robots, objects, locations, and task constraints. Through the prompt, the generated plan specifies robot assignments, actions, and dependencies between tasks. During execution, the frontend sends \emph{commitments} to the backend to report task progress and execution status, while the backend continuously updates the current plan across multiple interaction turns. When execution fails or the user issues a control instruction, the system can either directly modify the ongoing execution or generate a revised plan for the remaining tasks. Additional prompt details are provided in Appendix~\ref{app:prompt}.

\section{User Study}
We conducted a within-subjects study to evaluate how \oursystem supports users in understanding and intervening in LLM-based multi-robot coordination. We compared \oursystem with an LLM-based conversational interface through which users provided natural-language instructions and received textual feedback about the robot team. Both conditions used the same underlying planning model and robot capabilities. Specifically, we are interested in the following RQs:

\textbf{RQ1}: How does \oursystem affect users' performance and interaction effort when supervising and intervening in LLM-based multi-robot coordination, compared with a traditional LLM-based conversational interface?

\textbf{RQ2}: How do the two interfaces differ in users' situation awareness, perceived workload, control, transparency, and trust?

\textbf{RQ3}: How do users approach supervision and intervention with each interface, and what benefits and difficulties do they encounter?

\subsection{Participants and Settings}
This study was approved by the [University] Institutional Review Board (IRB). Participants were recruited through university email advertisements and offline posters. Each participant who completed the study received a \$30 gift card as compensation.
A total of 20 participants (6 female and 14 male) were included in the final analysis. Their ages ranged from 19 to 55 years (M = 26.4, SD = 7.72). Among them, seven participants were heavy VR users, 11 had used VR no more than three times, and two had never used VR before.
Regarding experience with AI tools, 12 participants reported being familiar with their use, four considered themselves experts, and four had only a basic understanding. For experience with robotic systems, three participants reported being very familiar with robotic systems and having worked with them in research or professional settings. Another four had occasionally operated or programmed robots in coursework, hobby projects, or at work. Four had only basic experience and had controlled a robotic system once. Seven had seen robotic systems but had never controlled one, while the remaining two reported no prior familiarity with robotic systems. Experience with MRS was more limited. Only one participant reported frequently using MRS in research or work. Four had occasionally controlled multiple robots at the same time, and seven had seen multiple robots operating together but had never controlled such a system. The remaining eight participants reported no prior knowledge of or experience with MRS. The study took place in a single room using Quest 3 with controller, where participants experienced the MR interface in a simulated environment presented through a VR headset. Participants remained seated in a swivel chair and could turn in place, but were not able to walk physically during the study for safety reasons. This setup was a practical compromise and we discuss this restriction for real-world MR use in the limitations section.

\subsection{Study Conditions}

We used a within-subjects design with two conditions: \oursystem(Fig.~\ref{fig:SD1-1}-1,2,3) and a conversational LLM-based interface(Fig.~\ref{fig:SD1-1}-4,5). Each participant completed one formal task(~\ref{sec:taskdesign}) with each interface. Both condition order and task-to-condition assignment were counterbalanced across participants.
We selected a conventional LLM-based conversational interface as the baseline because text-and-voice conversation is a common way of interacting with LLM agents and has also been adopted in immersive systems~\cite{li2025xr, izadi2025androidxr}. Both conditions used the same LLM planner, internal commitment representation, robot capabilities, and supported different intervention types (for more details, see Appendix~\ref{app:prompt}). In the conversational interface, planning decisions, requests for user input, and execution updates were presented in a chronological text stream; users specified both the target of an intervention and the intended change through natural language. \oursystem externalized the corresponding coordination state through persistent, linked representations across the minimap and situated MR view, and allowed users to scope interventions through spatial selection and context-specific controls.

\subsection{Task Design}\label{sec:taskdesign}

We designed two matched household tasks (Tasks A and B) to evaluate supervision and intervention during spatially distributed multi-robot coordination. Each task took place in a two-story house and involved cross-floor transport, off-screen robot activity, spatial references, and robot handoffs. The team comprised a fixed-base robotic arm, a wheeled delivery robot limited to one floor, and a drone capable of cross-floor transport~\cite{yu2026dynahmrc}. 

To cover different points at which human involvement may be needed, we identified eight intervention types from prior work on human--multi-robot interaction and LLM-based robot planning (details see Appendix~\ref{app:intervention_taxonomy}): information clarification, plan selection or optimization, authorization or access, task or goal updates, pre-failure action correction, failure recovery, capability-gap assistance, and contextual or norm adjustment. Each task contained nine planned intervention events The information-clarification event also included a conditional failure branch: selecting the wrong object caused the task to fail immediately and required the participant to select again. We treated this branch as part of the same event rather than as an additional intervention event. Thus, each participant encountered all eight intervention types under both conditions. Table~\ref{tab:intervention_types} summarizes Task A. 

We also distinguished intervention type from initiation mode, which specified who raised the need for intervention~\cite{hardin2009using,sellner2006coordinated}. To make events comparable across conditions, human-initiated events began with the same first-person contextual update presented through text and speech on a separate VR panel. The first-person framing presented the update as information known by the participant rather than as a system request. The update did not prescribe a response; participants raised the issue through the assigned interface. Agent-initiated events instead used predefined system requests, presented through a spatial label in \oursystem or a text message in the Conversational Interface. This design standardized event exposure but did not measure whether participants would detect and initiate these interventions spontaneously, as discussed in the Limitations.

Tasks A and B differed in layout, objects, destinations, and narrative content to reduce carryover, while keeping robot capabilities, intervention types, initiation modes, and dependency structures constant. Events occurred at corresponding dependency points, although robots could execute actions concurrently between events.

\begin{table*}[t]
\centering
\caption{Intervention events in Task A, including the messages presented to participants, affected robots, intervention types, and initiation modes. Human-initiated events were cued by condition-independent text-and-speech updates on a separate VR panel, whereas agent-initiated events requested input through a spatial label in \oursystem or a text message in the Conversational Interface.}
\label{tab:task_scenarios}

\scriptsize
\setlength{\tabcolsep}{3pt}
\renewcommand{\arraystretch}{1.18}

\begin{tabularx}{\textwidth}{
    >{\centering\arraybackslash}p{0.35cm}
    >{\raggedright\arraybackslash}p{2.0cm}
    >{\raggedright\arraybackslash}X
    >{\raggedright\arraybackslash}p{1.45cm}
    >{\raggedright\arraybackslash}p{2.5cm}
    >{\centering\arraybackslash}p{1.1cm}
}
\toprule

\textbf{\#} &
\textbf{Intervention Event} &
\textbf{Message Presented to Participant} &
\textbf{Robot} &
\textbf{Intervention Type} &
\textbf{Initiator}
\\

\midrule

1 &
Ambiguous object reference &
Which item do you want to choose? Select one highlighted option in the space [Glasses 1] [Glasses 2] &
Robot Arm &
Information Clarification &
Agent
\\

\addlinespace[2pt]

2 &
Route preference &
Choose a route preference: Default route is safer but slower; the Fast route is quicker but riskier. &
Drone &
Plan Selection or Optimization &
Agent
\\

\addlinespace[2pt]

3 &
Potentially unstable grasp &
Oops! I noticed that the robotic arm is pinching the glasses by the temple, so they may slip. I need to adjust the grasp so it holds the frame securely. &
Robot Team &
Pre-Failure Action Correction &
Human
\\

\addlinespace[2pt]

4 &
Delivery robot blocked &
Execution failure: The robot is blocked. Assign another robot to take over or roll the delivery robot back to its previous checkpoint and replan a route. &
Delivery Robot &
Failure Recovery &
Agent
\\

\addlinespace[2pt]

5 &
Private-room access &
The spaceguest room is marked PRIVATE. Allow the drone to enter? &
Drone &
Authorization or Access &
Agent
\\

\addlinespace[2pt]

6 &
Recipient relocation &
I just realized that the Jim has moved to the studio. I have to pick the glasses back up from the bedside table and deliver them to the desk in the studio. &
Robot Team &
Task or Goal Update &
Human
\\

\addlinespace[2pt]

7 &
Closed-door assistance &
Please open the door to allow the drone to enter. &
Drone &
Capability-Gap Assistance &
Agent
\\

\addlinespace[2pt]

8 &
Do-not-disturb preference &
I just realized that the Jim doesn't want to be disturbed while working. I need to place the item on the bedside table instead. &
Robot Team &
Contextual or Norm Adjustment &
Human
\\

\addlinespace[2pt]

9 &
Drone localization error &
Drone localization error --- A localization error made the drone drift off target --- the item is being placed on the edge of the bedside table, where it may fall. Please instruct the drone to move the item to a safe spot. &
Drone &
Failure Recovery &
Agent
\\

\bottomrule
\end{tabularx}
\end{table*}

\subsection{Study Procedure}
The study lasted approximately two hours. Participants first signed an informed consent form and completed a demographic questionnaire. The researcher then provided a standardized orientation lasting approximately 30 minutes, covering the study procedure, the multi-robot task, the robots' capabilities, and the interaction techniques available in both conditions. 
Participants then put on the headset and entered the tutorial task to learn how to operate in two conditions. They practiced monitoring the robot team, initiating interventions, and responding to narrative-cued and agent-prompted events. The tutorial session finished after participants confirmed that they understood the task and could use both interfaces. Participants then took a 5--10-minute break from VR and proceeded only after confirming that they felt able to continue without substantial discomfort or cybersickness. 

Each participant completed two trials, one task with each interface. Participants were assigned to one of four sequences that counterbalanced interface order and task--interface pairing. Each trial lasted approximately 15--30 minutes. During each trial, the system logged intervention completion times and participants' prompt numbers and lengths. Immediately after each trial, participants completed questionnaires measuring workload, situation awareness, trust, transparency, and perceived control. After both trials, participants completed an approximately 30-minute semi-structured interview about their interaction strategies, experiences with each interface, interface preferences, and perceived benefits and limitations. Section~\ref{DC} describes these measures and the interview protocol in detail.

\subsection{Data Collection}\label{DC}

\paragraph{System Logs (RQ1)}
During each formal task, the system recorded timestamped intervention events, natural-language prompts, and backend requests and responses. From these logs, we derived three measures: intervention completion time, prompt count, and mean prompt length. Completion time and interaction effort are commonly used measures for evaluating human supervision of multi-robot systems~\cite{olsen2003metrics, steinfeld2006common}. Prompt number and length reflect the effort required to express intent through natural language. This cost is inherent in language-based interaction~\cite{clark1991grounding} and can increase further when users instruct an LLM~\cite{zamfirescu2023johnny}. Intervention completion time was defined as the elapsed time from the onset of an intervention event until the participant completed the required response. Prompt count was the number of prompts issued within a task, and mean prompt length was the average number of words per prompt. 

\paragraph{Questionnaires (RQ2)}
After each formal task, participants completed questionnaires measuring workload, situation awareness, trust, transparency, and perceived control. These measures were selected because situation awareness and workload are central to supervising multiple robots simultaneously~\cite{endsley1995toward, chen2014human}. Trust and transparency affect whether users can appropriately rely on autonomous systems~\cite{lee2004trust}, while perceived control is also important when automation changes how users intervene in robots and reassign tasks~\cite{parasuraman2000model, shneiderman2020human}. We measured workload using the Raw NASA-TLX~\cite{hart1988development} and situation awareness using SART 10D~\cite{selcon1990evaluation}. Trust was measured using MDMT v1~\cite{malle2023measuring}, from which we calculated Capacity Trust and Moral Trust scores. Transparency was measured using the Explainability and Predictability subscales of TOROS~\cite{angelopoulos2025measuring}. Perceived control was measured using three 7-point items developed for this study. Appendix~\ref{app:CQ} provides the questionnaire items, and Section~\ref{sec:data-analysis} describes scoring and reliability analyses.

\paragraph{Semi-structured Interviews (RQ3)}
After completing both tasks, participants took part in an approximately 30-minute semi-structured interview. We asked them to compare the two interfaces in how they monitored the robot team, maintained awareness, referred to spatial information, made intervention decisions, and controlled or reassigned robots. Follow-up questions elicited specific examples of their strategies, difficulties, and responses to system behavior. Participants also explained which interface they preferred and reflected on the benefits and limitations of each interface and its features. Interviews were audio-recorded and transcribed for qualitative analysis.

\subsection{Data Analysis}\label{sec:data-analysis}
For the logging data, two tasks presented the same eight intervention types in different orders. Because events of different types were not directly comparable, we paired each event in Task A with the same-type event in Task B, yielding one Baseline and one MRPilot value per intervention type. Failure Recovery occurred twice in each task, so its completion time under each condition was the sum of the two events. Prompt number was the number of instructions issued to the system within a trial, and prompt length was the mean number of words per prompt.

For the questionnaire data, we did not apply the pairwise weighting procedure for NASA-TLX and therefore report Raw TLX, the unweighted mean of the six subscales~\cite{hart1988development}, as the overall workload score alongside the individual subscales. The Performance scale was anchored at perfect (0) and failure (100), so higher scores on all six subscales indicate higher workload. SART 10D was calculated from the sums of its three domains using the formula $SA = Understanding - (Demand - Supply)$. MDMT v1 was scored from 0 to 7 separately for each subscale; the explicit Does Not Fit option was treated as missing, and the participant was excluded only from the analysis of that outcome. TOROS used its 1–7 scale, with the two negatively worded items reverse-scored. Perceived control was measured with three self-designed items on a 1–7 scale. We first assessed reliability using Cronbach's $\alpha$ with Feldt 95\% confidence intervals. The prespecified criteria were $\alpha \geq .70$ in both conditions and corrected item total correlations $\geq .30$. The scale met these criteria under both conditions, so we analyzed the mean of the three items as a single outcome.

For the system logging data and NASA-TLX results, we first assessed the normality of paired differences with the Shapiro–Wilk test at the ($\alpha=.05$ level), then used a paired-samples t-test when differences were normally distributed and a Wilcoxon signed-rank test otherwise. Individual Likert-type items within questionnaire subscales were analyzed with the Wilcoxon signed-rank test because of their ordinal nature.

For the qualitative data, interview transcripts and audio recorded during the task sessions were analyzed using reflexive thematic analysis~\cite{braun2006using}, combining deductive and inductive coding. We developed an initial set of deductive codes around the study's core constructs and, in parallel, conducted open inductive coding to capture patterns beyond them. The analysis proceeded iteratively through familiarization, initial coding, candidate theme development, checking and revising themes against the transcripts, and defining and naming the final themes. The first author conducted the coding, and the emerging themes were reviewed and refined through multiple rounds of discussion with the research team.

\section{Results}

Our results address the three research questions. For reporting, when the data were normally distributed, we report the mean (M) and standard deviation (SD) for each condition, together with the paired samples t test statistic, p value, and standardized effect size. When the data were not normally distributed, we report the median (Mdn) and interquartile range (IQR) for each condition, together with the Wilcoxon signed rank statistic, p value, and rank biserial correlation. In the following analyses, the subscripts $b$ and $m$ denote Baseline and \oursystem, respectively.

\begin{figure*}[t]
    \centering
    \includegraphics[width=0.8\textwidth]{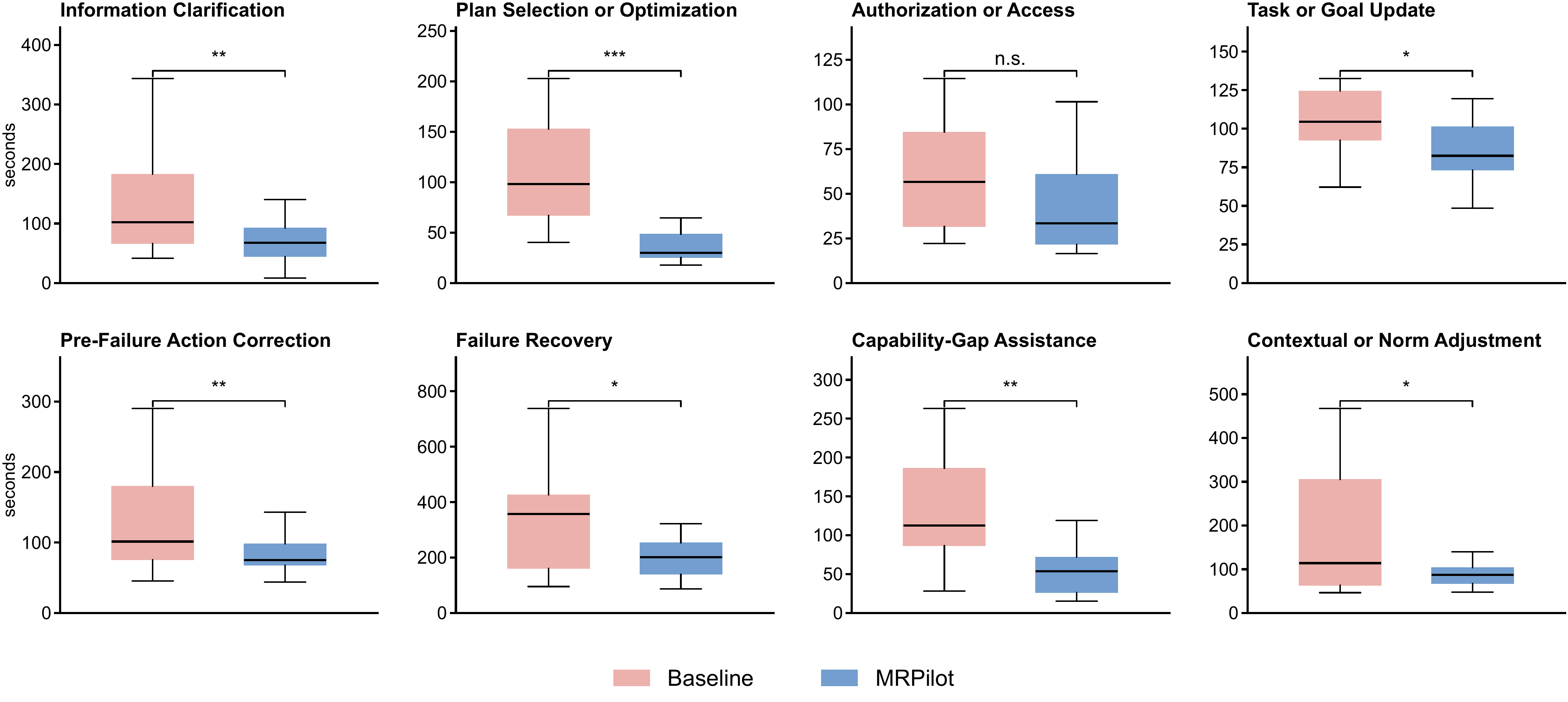}
    \caption{The Results of Intervention Type Complete Time, Note: Throughout all figures, significance levels are indicated as follows: * $p<.05$, ** $p<.01$, *** $p<.001$}
    \label{fig:intervention_time}
\end{figure*}

\begin{figure*}[t]
    \centering
    \includegraphics[width=0.55\textwidth]{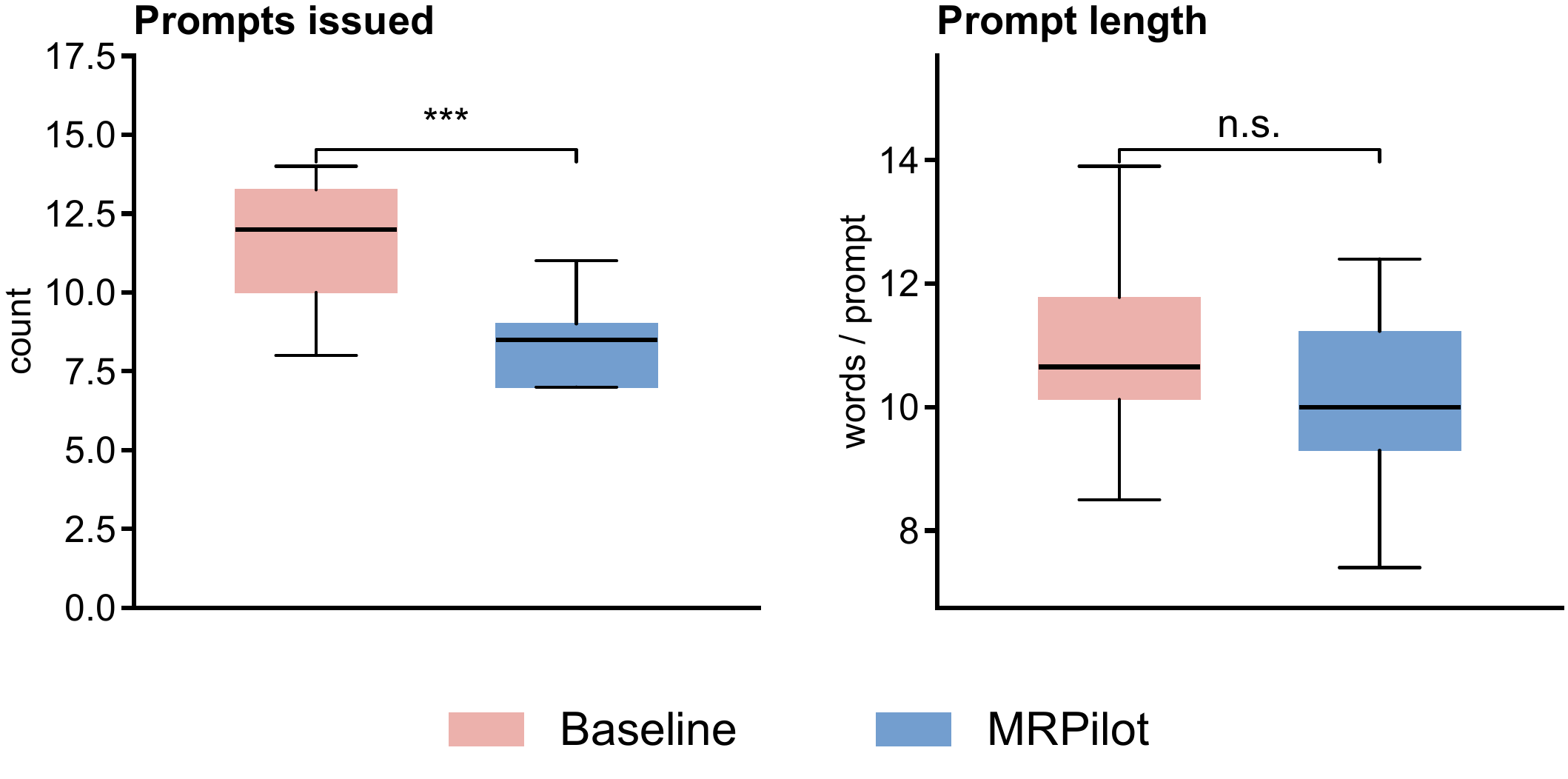}
    \caption{The Result of Prompt issued and Prompt Length}
    \label{fig:interactioncost}
\end{figure*}

\subsection{Users' Performance and Interaction Effort (RQ1)}\label{UPIE}
For each outcome variable, we first used the Shapiro–Wilk test to assess normality. Prompt Number and the completion times of three types of intervention events were normally distributed: Plan Selection or Optimization, Authorization or Access, and Capability-Gap Assistance. The remaining data were not normally distributed.

We next compared \oursystem and Baseline in completion time across different intervention types. \oursystem resulted in shorter completion times than Baseline for seven types of intervention events (Fig.~\ref{fig:intervention_time}). Tasks that could be resolved directly through system-supported interactions were completed the fastest, including Information Clarification (Type 1; $Mdn_b = 102.24$, $IQR [66.40, 182.82]$, $Mdn_m = 67.49$, $IQR [44.37, 92.33]$; $W = 23$, $p < .01$, $r = -0.78$), Plan Selection or Optimization (Type 2; $M_b = 110.58$, $SD = 51.08$, $M_m = 37.01$, $SD = 18.48$; $t(19) = -7.19$, $p < .001$, $d_z = -1.61$), and Failure Recovery (Type 6; $Mdn_b = 356.87$, $IQR [161.13, 426.20]$; $Mdn_m = 201.79$, $IQR [139.87, 252.84]$; $W = 49$, $p = 0.036$, $r = -0.53$). Similar results were found for Task or Goal Update (Type 4; $Mdn_b = 104.44$ s, $IQR [92.61, 124.45]$; $Mdn_m = 82.45$ s, $IQR [73.21, 101.18]$; $W = 41$, $p = 0.015$, $r = -0.61$) and Contextual or Norm Adjustment (Type 8; $Mdn_b = 114.01$ s, $IQR [63.60, 305.11]$; $Mdn_m = 87.30$ s, $IQR [67.66, 103.56]$; $W = 50$, $p = 0.040$, $r = -0.52$).

For tasks requiring more spatial judgment, such as interactions across floors or over longer distances, \oursystem also improved interaction efficiency. This was observed in Pre-Failure Action Correction (Type 5; $Mdn_b = 101.40$ s, $IQR [75.68, 179.82]$; $Mdn_m = 75.25$ s, $IQR [67.86, 98.10]$; $W = 25$, $p = 0.002$, $r = -0.76$) and Failure and Capability-Gap Assistance (Type 7; $M_b = 143.98$ s, $SD = 98.96$; $M_m = 59.94$ s, $SD = 45.31$; $t(19) = -3.44$, $p = 0.003$, $d_z = -0.77$). In contrast, for tasks occurring on the same floor or involving only short-distance interactions, there was no significant difference between \oursystem and Baseline. This was observed in Authorization or Access (Type 3; $Mdn_b = 64.38$, $SD = 39.40$, $Mdn_m = 46.68$, $SD = 35.22$; $t(19) = -1.31$, $p = .206$, $d_z = -0.29$).

For interaction effort (Fig~\ref{fig:interactioncost}), there was no significant difference between conditions in average prompt length ($Mdn_b = 10.65$, $IQR [10.12, 11.78]$. $Mdn_m = 10.00$, $IQR [9.30, 11.22]$; $W = 53.5$, $p = 0.095$, $r = -0.44$). However, participants issued fewer prompts with \oursystem, with an average reduction of 4.00 prompts ($M_m = 8.40$, $SD = 1.27$, $M_b = 12.40$, $SD = 3.66$; $t(19) = -4.89$, $p <.001$, $d_z = -1.09$). Thus, the reduction in interaction cost with \oursystem came from fewer interaction turns rather than shorter interaction content.

\subsection{Users' Situation Awareness, Perceived Workload, Control, Transparency, and Trust(RQ2)}

\subsubsection{Perceived Workload}

\begin{figure*}[t]
    \centering
    \includegraphics[width=0.8\textwidth]{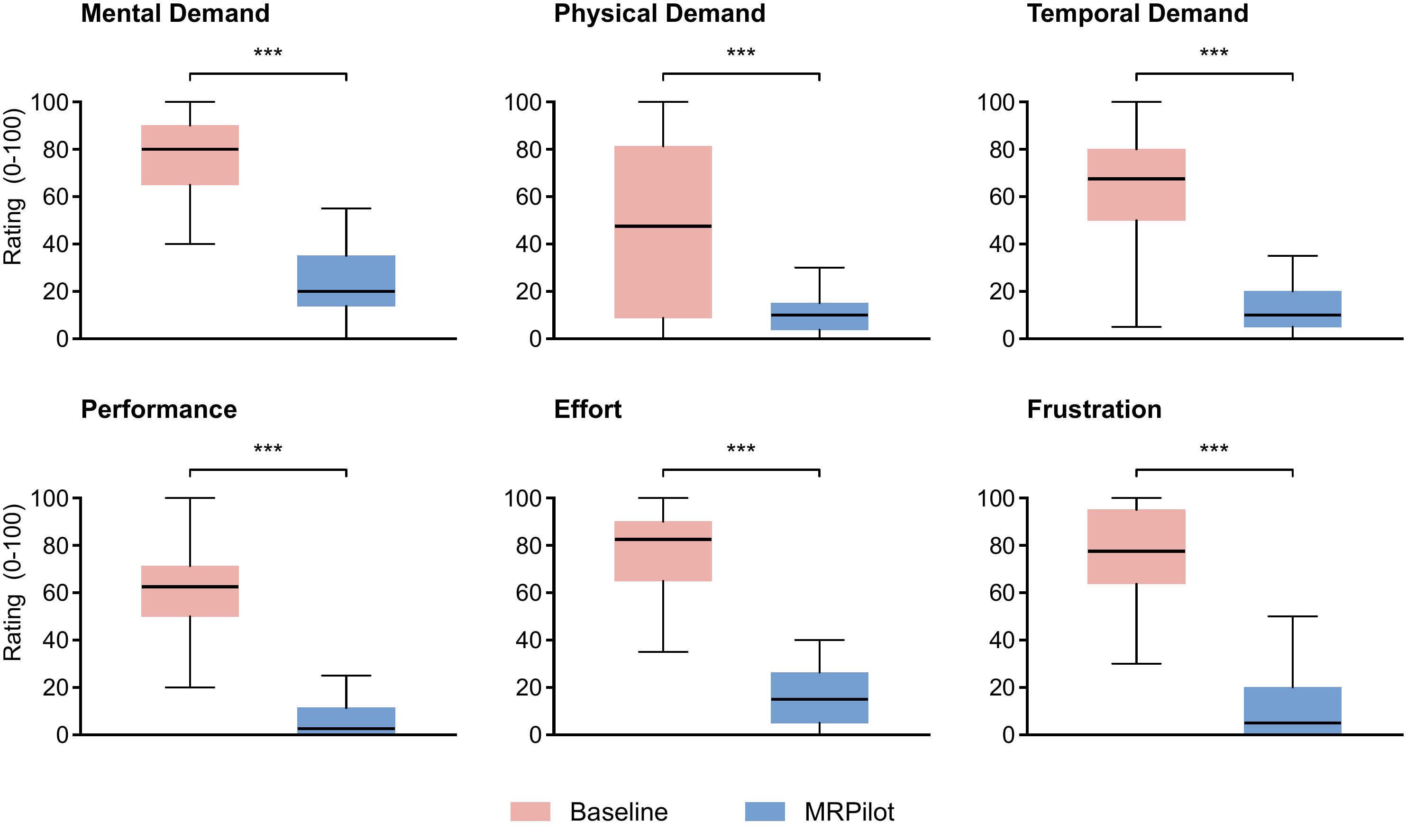}
    \caption{NASA TLX Questionnaire Results}
    \label{fig:nasa-tlx}
\end{figure*}
To compare subjective workload across conditions, we first used the Shapiro–Wilk test to assess the normality of the within-participant differences for each NASA-TLX outcome. \oursystem produced significantly lower scores than Baseline on all six subscales (Fig~\ref{fig:nasa-tlx}). The within-participant differences for five subscales were consistent with normality. Paired-samples t-tests showed significant reductions in Mental Demand ($M_b = 75.25$, $M_m = 23.00$; $t(19) = -10.28$, $p < .001$, $d_z = -2.30$), Temporal Demand ($M_b = 65.00$, $M_m = 14.25$; $t(19) = -7.68$, $p < .001$, $d_z = -1.72$), Performance ($M_b = 60.25$, $M_m = 8.75$; $t(19) = -8.25$, $p < .001$, $d_z = -1.84$), Effort ($M_b = 75.50$, $M_m = 16.25$; $t(19) = -9.54$, $p < .001$, $d_z = -2.13$), and Frustration ($M_b = 72.75$, $M_m = 13.5$; $t(19) = -8.93$, $p < .001$, $d_z = -2.00$). The within-participant differences for Physical Demand departed from normality. Physical Demand was also significantly lower with \oursystem ($Mdn_b = 47.5$, $Mdn_m = 10.0$; $W = 3$, $Z = -3.48$, $p < .001$, $r = -0.96$).

Physical Demand showed the smallest reduction, at approximately 35 points. Its scores under Baseline were also substantially more dispersed than those of the other subscales. These results show that the interaction provided by \oursystem effectively reduced subjective workload. The smaller change in Physical Demand may reflect the study setting, as the tasks were simulated in a virtual reality environment and imposed relatively limited physical activity.

\subsubsection{Situation Awareness}
For discrete outcomes, we used nonparametric statistical methods. \oursystem significantly improved scores across all three SART dimensions compared with Baseline (Figure~\ref{fig:questionnaire-results}). Overall SA was significantly higher with \oursystem ($Mdn_b = 7.00$, $IQR [4.75, 13.25]$, $Mdn_m = 33.50$, $IQR [29.00, 40.00]$; $W = 3$, $p < .001$, $r_{rb} = .97$). Attentional Demand decreased ($Mdn_b = 14.00$, $Mdn_m = 5.00$; $W = 0$, $p < .001$, $r_{rb} = -1.00$), while Attentional Supply increased ($Mdn_b = 14.50$, $Mdn_m = 20.50$; $W = 21$, $p < .01$, $r_{rb} = .80$), as did Understanding of the Situation ($Mdn_b = 8.50$, $Mdn_m = 18.50$; $W = 6.5$, $p < .001$, $r_{rb} = .94$). These results indicate that the interaction provided by \oursystem reduced the attentional effort required for operators to maintain situation awareness while helping them develop a more complete understanding of the robot team's state.

\subsubsection{Trust}

\oursystem was significantly higher than in Baseline (Fig.~\ref{fig:questionnaire-results}). More specifically, the increase was larger for Capacity Trust ($Mdn_b = 2.94$, $Mdn_m = 6.75$; $W = 2$, $p < .001$, $r_{rb} = .98$), whereas the increase in Moral Trust was relatively smaller ($Mdn_b = 4.71$, $Mdn_m = 6.75$; $W = 0$, $p = 0.001$,$r_{rb} = 1.00$). In addition, participants selected Does Not Fit for the moral adjectives less often with \oursystem than with Baseline (38 vs. 49 responses). Taken together, these results suggest that the increase in trust was driven primarily by participants' confidence in the capabilities provided by \oursystem. When the system’s capabilities were directly reflected in the spatial task, participants’ trust in the system also increased.

\subsubsection{Transparency}
\begin{figure}
    \centering
    \includegraphics[width=0.8\linewidth]{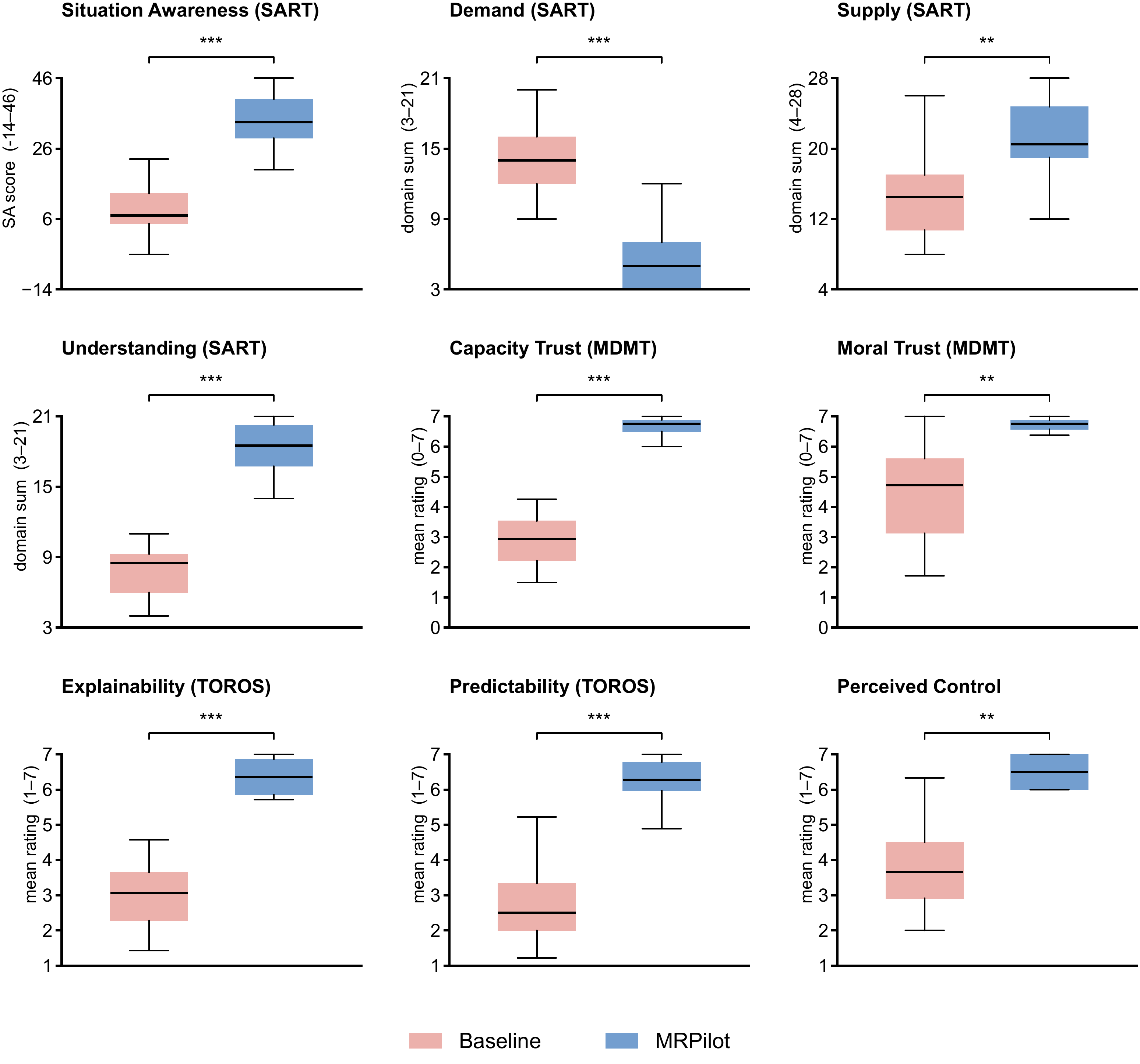}
    \caption{The Results of SART, MDMT v1, TOROS and self report questionnaires}
    \label{fig:questionnaire-results}
\end{figure}

\oursystem also significantly improved the transparency of robot operation compared with Baseline (Fig~\ref{fig:questionnaire-results}. This was reflected in both Explainability ($Mdn_b = 3.07$, $Mdn_m = 6.36$; $W = 0$, $p < .001$, $r_{rb} = 1.00$) and Predictability ($Mdn_b = 2.50$, $Mdn_m = 6.28$; $W = 1.5$, $p < .001$, $r_{rb} = .99$). These results indicate that \oursystem improved the transparency of the robot team by making robot actions easier to explain and allowing users to preview planned actions before execution.

\subsubsection{Perceived Controllability}
\oursystem significantly increased participants' perceived control over the robot team compared with Baseline ($Mdn_b = 3.67$, $IQR [2.92, 4.50]$, $Mdn_m = 6.50$, $IQR [6.00, 7.00]$; $W = 16.5$, $p < .01$, $r_{rb} = .81$). These results suggest that the direct intervention mechanisms provided by \oursystem strengthened operators' sense of control, whereas the text and voice based interaction in Baseline provided less direct means for intervention.

\subsection{Benefits and Difficulties for Supervision and Intervention(RQ3)}
\subsubsection{Direct control increases users' autonomy in intervention, but introduces the cost for using the available functions}
Some participants reported that they felt more in control when using \oursystem, while the Baseline felt more system-driven. P14 stated, \textit{``In the base I feel like the system is in control, but in the AR environment, I feel like I'm the one in control.''} In the Baseline, all interventions had to be completed through voice prompts. Users needed to fully specify the instruction, including the robot, the robot’s location, and the target location. If any detail was missing, they had to enter the prompt again, and modifying the plan meant restarting the entire plan (P4). In contrast, in \oursystem, participants could directly select a robot and operate it. P9 said, \textit{``can just say do this as opposed to ... get the drone to do this,''} and used the revert, delete, and path buttons to achieve greater autonomy. P4 found it \textit{``easier to ... revert it to its original position [because] you wouldn't have to specify the prompt.''}

This autonomy also came with a set of functions and control methods that participants needed to learn. P13 said \textit{``there's a lot of parts to it''} although \textit{``once you figure out the system it makes it pretty easy.''} The purpose of different controls was not always clear. After recording a command, P11 faced two similar buttons and said, \textit{``just didn't know which one is the execution''} The controls themselves also required practice because users needed to calibrate their interactions each time they issued a command (P14, P16). The Baseline did not have these issues. P17 thought it was \textit{``pretty clear that [you] just open the control panel and talk to it''} P16 preferred \textit{``just one central control panel ... I don't have to keep switching.''} and thought this approach \textit{``a bit more convenient''}when there are more robots.

\subsubsection{Spatial visualization enables faster localization and improves situation awareness, but is constrained by the field of view}
In the Baseline, users had poorer awareness of the robots' spatial locations. P9 noted that the failure alert \textit{``didn't tell me the location it just told me this is the status do something about it.''} P1 also said that they were still \textit{``uncertain about ... what is actually failing and what could be the right solution.''} In contrast, \oursystem made localization direct. Robot labels, route lines, and off-screen cues showed where each robot was and where a failure had occurred, even when the robot was behind a wall or on another floor. P4 could follow \textit{``a line pointing towards the robot''} and P6 said \textit{``can directly see even though the robot isn't on my layer or behind the wall.''} Participants could therefore supervise the robot team without moving from where they stood.

However, the increase in situation awareness provided by these overlays was still limited by what participants could see. When a robot was on another floor, the system could show that the robot was there, but not what was happening around it. When P14 was upstairs and received a prompt saying \textit{``a child needs to slow down but can't see if there really is a child or what's going on.''} Whether an object had been placed successfully or whether a hazard was actually present was the information that determined whether users needed to intervene. In both interfaces, this information could only be communicated to participants through the robots’ reports. The Baseline had an advantage in this respect. Because participants already had to walk to the robot’s location themselves, the local situation would naturally enter their field of view. As P14 said \textit{``you really have to observe them with your own eyes.''}

\subsubsection{Visualizing the execution process increases user trust, but reduces the explainability provided through text}
The Baseline was described as an opaque black box, which limited participants’ trust in the system. P11 said \textit{``It's like you talk to a black box''}; P5 said it was \textit{``not fully trusted because the text sometimes can miss some information.''} Participants therefore compensated for this uncertainty through additional checking. P9 had \textit{``to have that panel open in order to see the status update.''}In \oursystem, trust came from being able to see the execution process itself. P11 said \textit{``in between, you don't have a block, there's no blank space, you always know what it's doing, so it builds trust.''} P5 also chose \oursystem because \textit{``there are a lot of features I can see I can trust.''} As a result, the need for continuous monitoring was reduced. P3 said they could \textit{``focus on the high level decision.''}

However, the visualization did not provide explanatory information. The Baseline presented the cause of a problem and the required action in text, which participants could read directly. P12 knew that a door was locked \textit{``because without that message I would assume that it got to its destination''}; P17 also saw an advantage in that \textit{``does let you know that if something is broken down and you have to change it.''} In \oursystem, the same information was conveyed through icons, labels, and lines. These elements could show what happened and where it happened, but not why, and they could also be overlooked. P8 said they \textit{``know what's the difference between the two choices.''} while P11 wanted to understand \textit{``the underlying logic of the route.''} Thus, trust in the Baseline was based on receiving the correct result, while in \oursystem, users trusted the visualized execution process but still had to infer the logic behind it.

\section{Discussion}

\subsection{Reflection on the Value of Spatial Supervision and Intervention}

RQ1 helps identify the interaction demands for which \oursystem was most useful. \oursystem reduced completion time for seven intervention types, with the largest gains in tasks involving cross-floor interaction, longer-distance spatial reference, or direct manipulation; the two interfaces performed similarly for \textit{Authorization or Access}. \oursystem also reduced the number of prompts without shortening individual prompts. Its advantage was therefore conditional: it was most pronounced when participants faced complex spatial relationships, such as locating a robot, identifying a plan element, or making a spatially grounded correction, whereas for update tasks with simpler descriptions and limited spatial complexity the conversational interface performed comparably.

The RQ2 results indicate that \oursystem improved situation awareness, perceived control, transparency, and trust while reducing workload. By externalizing robot locations, plans, and intervention targets, its spatial views and direct controls made distributed task state easier to monitor and act on. These benefits reflect the fit between \oursystem and the spatially distributed tasks examined here, and our findings are bounded accordingly. The two-story environments and site-specific interventions created repeated demands for spatial monitoring and situated action, so the results apply primarily to spatially distributed MRS tasks rather than to tasks conducted in a single visible area or requiring little spatial reference. Moreover, because \oursystem combined MR visualization, plan inspection, and direct controls, the study evaluates this integrated configuration rather than isolating the contribution of each component.

\subsection{Design Implications for Human--MRS Interaction}

\subsubsection{Supporting Coordination-Aware Intervention Across Multiple Scales}
MRPilot allowed participants to intervene directly in an individual robot or plan component, but actions in an MRS remain connected through waits, handoffs, shared resources, and downstream \emph{commitments}. Interface design should therefore connect local intervention with team-level coordination. When a user modifies one robot's action, these dependencies can be represented through temporal networks with spatial and resource constraints~\cite{gombolay2018fast, wang2022heterogeneous} to identify which concurrent and downstream actions are affected. The system should preserve \emph{commitments} that remain valid rather than rescheduling the entire robot team, while communicating the scope of the resulting plan changes to the user. This would extend direct spatial manipulation from editing an isolated action to managing its downstream effects across the team, which may improve team performance and trust~\cite{natarajan2024mixed}. Future work should also develop coordination-aware partial replanning mechanisms that combine spatial editing with such schedulers and examine how they operate when multiple robots execute tasks concurrently, since task durations may be stochastic and change over time~\cite{liu2021coordinating, altundas2022learning}. Evaluation should examine whether users understand the downstream effects of an intervention, whether the modified plan remains internally consistent, whether changes unintentionally affect other robots~\cite{adekoya2026multi}, and how these outcomes vary with robot team size and dependency density.

\subsubsection{Adapting Information and Initiative to Supervisory Need}
MRPilot improved localization across floors, but participants still lacked detailed information about conditions around robots outside their field of view. The qualitative results also identified costs associated with learning different control methods and switching between them. The appropriate amount of information and level of human involvement should therefore depend on the current supervisory need~\cite{parasuraman2000model, sheridan2011adaptive}. The interface should coordinate information detail and interaction initiative based on task phase, urgency, uncertainty, potential consequences, robot visibility, and the user's preferred level of automation to support adaptive information presentation~\cite{chiou2021mixed, rule2012designing}. For routine and high-confidence execution, peripheral awareness information may be sufficient. For uncertain or consequential decisions, the system may need to provide more detailed evidence, request user confirmation, or support direct intervention~\cite{fong2003multi, hardin2009using}. Such strategies should remain predictable, inspectable, and adjustable by users~\cite{chen2014human, methnani2024s}. Future work should vary both task conditions and user responsibilities, and examine missed intervention opportunities, unnecessary interruptions, workload, mode confusion, and intervention quality~\cite{kaber2004effects, ramesh2023experimental}. Because interaction initiation modes were not manipulated in our study, this design implication requires further validation.

\subsubsection{Moving from Visible Execution to Verifiable and Calibrated Reliance}
Participants rated MRPilot as more transparent and trustworthy, although both conditions used the same planner and participants still had difficulty understanding why some routes or decisions were selected~\cite{chakraborti2020emerging}. Future systems should distinguish among observing an action, receiving an explanation for it, and verifying that it satisfies task and safety constraints~\cite{luckcuck2019formal}. Spatial representations should connect visible plans to evidence about robot capabilities, preconditions, dependencies, sensed conditions, and uncertainty~\cite{kunze2019automation}, while distinguishing observed state from model inference and verified information~\cite{chen2018situation}. The goal is calibrated reliance, in which users accept well-supported actions and question actions whose evidence is incomplete~\cite{wang2016trust}. Future work should vary planner accuracy and explanation reliability, then measure error detection, appropriate acceptance or rejection, intervention decisions, and users' ability to distinguish plausible explanations from verified accounts~\cite{hoffman2023measures}. Such evaluations would determine whether perceived transparency corresponds to an accurate understanding of robot behavior and its constraints~\cite{chromik2021think}.

\subsection{Limitations and Future Work}
Our evaluation used a fixed robot team and a predefined task set, which ensured fair comparison but limits generalizability. Supervisory capacity does not scale linearly with team size \cite{olsen2004fan}, and heterogeneous teams impose different demands than homogeneous ones \cite{chen2014human}. The advantages of \oursystem may therefore change with larger or more heterogeneous teams, or with longer, open-ended tasks. Future work should vary team composition and test the system over longer periods with dynamically generated tasks, such as search and rescue.

Our study also compared the MR interface and the conversational Baseline as complete systems. The MR condition added situated 3D visualization of robot states, spatial reference to locations and goals, and real-time embodied interaction. Our results thus show the overall advantage of spatially situated supervision over a text-centered interface, but not the contribution of each feature. Such holistic evaluation suits novel systems whose value lies in combining features \cite{greenberg2008usability}. Ablation or factorial studies could separate these effects, for example by adding a 2D map to the Baseline or disabling spatial anchoring and gesture input in MR, guided by existing VAM-HRI taxonomies \cite{walker2023virtual, suzuki2022augmented}.

The study was conducted in VR with seated participants. This ensured safety and consistency but omitted real-world sensing errors, motion uncertainty, communication latency, physical risks, occlusion, and natural operator movement. These factors could misalign overlays, delay feedback, or force operators to split attention between the robots and their own safety, so our results may not transfer directly to physical deployments. Future work should test the system with physical robots in real environments where operators can move freely.

Intervention opportunities were predefined so that all participants faced the same situations. This let us compare how participants responded to intervention needs, but not whether they could detect unexpected problems on their own, as real deployments require. Future work could introduce unscripted or randomly timed failures and measure whether and how quickly operators detect them.

To preserve realism, our tasks omitted uncommon combinations of intervention type and initiation mode. Because the two factors were not fully crossed, their effects cannot be separated, and our results do not extend to untested combinations. A fully crossed or targeted design could clarify how they interact.

Finally, we did not evaluate the underlying multi-robot planner, as our focus was on how operators supervise, verify, and intervene. System behavior thus depends on the LLM, whose outputs remain stochastic despite a fixed temperature and constrained prompts, so the same input may occasionally yield different task interpretations or assignments. Future work could pair our interface with multi-robot LLM planners \cite{kannan2024smart, mandi2024roco, chen2024scalable}, with hybrid pipelines that translate instructions into formal problems for classical planners with stronger correctness guarantees \cite{liu2023llm+}.

\section{Conclusion}
We present \oursystem, a mixed reality system that enables non-expert users to supervise and intervene in LLM-based MRS. Informed by design goals derived from a formative study with 12 non-expert users, \oursystem centers supervision on structured \emph{commitments} and situates it in spatial context. Before execution, users can inspect how their instructions are translated into coordinated plans, maintain awareness of distributed execution through synchronized situated and overview views, and repair individual robots or subtasks in context without reconstructing the entire task. A within-subjects study with 20 participants showed that, compared with a conversational interface using the same planning model and robot capabilities, \oursystem significantly improved situation awareness, sense of control, transparency, and trust while reducing workload and shortening completion times for spatially complex tasks such as cross-floor delivery and multi-robot collaboration. These improvements mainly came from fewer and more targeted interactions rather than shorter instructions. Continuous spatially situated visualization allowed participants to keep track of what the robot team was doing and where it was happening, while intuitive interaction mechanisms allowed them to act directly on that information. Based on these findings, we further summarize design implications for future supervision interfaces, including spatial reference through multimodal input, proactive adjustment of information detail based on real-time context, and embedding explainability into the spatial elements users interact with. As LLM-based MRS increasingly enter everyday environments, our work offers new design directions for more efficient and transparent interaction with such systems.

\bibliographystyle{ACM-Reference-Format}
\bibliography{sample-base}

\appendix

\section{Controllability Questionnaire}\label{app:CQ}
Participants rated each statement on a 7-point Likert scale,
ranging from 1 (\textit{Strongly disagree}) to
7 (\textit{Strongly agree}).

\begin{table*}[h]
\centering
\small

\renewcommand{\arraystretch}{1.8}
\setlength{\tabcolsep}{3pt}

\begin{tabularx}{\textwidth}{
    >{\raggedright\arraybackslash}X
    *{7}{>{\centering\arraybackslash}p{1.4cm}}
}

&
\textbf{1} &
\textbf{2} &
\textbf{3} &
\textbf{4} &
\textbf{5} &
\textbf{6} &
\textbf{7}
\\[-1mm]

&
\scriptsize Strongly disagree &
\scriptsize Disagree &
\scriptsize Somewhat disagree &
\scriptsize Neither agree nor disagree &
\scriptsize Somewhat agree &
\scriptsize Agree &
\scriptsize Strongly agree
\\

\midrule

I was able to make the robot team do what I intended.
&
$\bigcirc$ &
$\bigcirc$ &
$\bigcirc$ &
$\bigcirc$ &
$\bigcirc$ &
$\bigcirc$ &
$\bigcirc$
\\

When I wanted to change the robot team's plan, I was able to change it.
&
$\bigcirc$ &
$\bigcirc$ &
$\bigcirc$ &
$\bigcirc$ &
$\bigcirc$ &
$\bigcirc$ &
$\bigcirc$
\\

I was in control of how the task was carried out.
&
$\bigcirc$ &
$\bigcirc$ &
$\bigcirc$ &
$\bigcirc$ &
$\bigcirc$ &
$\bigcirc$ &
$\bigcirc$
\\

\bottomrule

\end{tabularx}
\end{table*}

\section{Prompt for the LLM Processor in \oursystem}\label{app:prompt}
\subsection{Spatial Information}
\subsubsection{Task A}
The following prompt defines the world state provided to the LLM
planner for Task A. The content is reproduced as used in the
study; line wrapping has been adjusted for readability.

\begin{promptbox}
## Scenario: task_a

### World State

Objects:
  - reading glass (at dining_table): a pair of reading glasses on the dining table, next to the robot_arm -- small and lightweight, requires robot_arm to pick up
  - sun glass (at living_room): a pair of sunglasses on the wine cabinet in the dining area of the living room -- small and lightweight

Locations:
  - dining_table: the dining table in the second-floor living room -- the robot_arm is fixed to it (tabletop height)
  - living_room: the second-floor living room, where the resident is; the delivery_robot and drone start here
  - drone_dock: the drone's docking station in the second-floor living room -- the designated handoff point where the drone lifts objects off the delivery_robot before flying up through the open stairwell to the third floor
  - corridor_third_floor: the third-floor corridor connecting the guest room and the studio (reachable only by drone)
  - guest_room: the first room off the third-floor corridor -- the guest room marked PRIVATE, the resident's resting area (bed + bedside table); entering requires the resident's privacy permission
  - bedside_table_guest_room: the bedside table next to the bed inside the guest room (third floor)
  - studio: the studio at the end of the third-floor corridor -- the resident's working area; its door is unlocked but kept closed (the resident opens it by hand)
  - bedside_table_studio: the bedside table inside the studio (third floor)
  - bedside_table_studio_edge: the near edge of the bedside table inside the studio (third floor) -- a precarious drop spot used by the system, not a user-facing destination
  - desk_studio: the desk inside the studio (third floor) -- the resident's work surface

Available robots:
  - robot_arm
  - delivery_robot
  - drone

Constraints:
  - the environment has two floors: the second floor (living room with the dining table) and the third floor (corridor with the guest room and studio), connected by an open-tread staircase through an open stairwell
  - robot_arm is fixed to the dining_table -- it can only pick up objects on the dining table and cannot move
  - delivery_robot operates ONLY on the second floor -- it cannot climb the staircase or reach any third-floor location
  - drone is the only robot that can reach the third floor -- it flies up through the open stairwell (cross-floor transport)
  - objects on the dining table must be picked up by robot_arm first
  - the guest_room is marked PRIVATE -- entering it requires the resident's privacy permission, which is requested at execution time; plan deliveries into it normally
  - a delivery into the guest_room is placed on its bedside table -- when the user asks to send an object to the guest room without naming a surface, the final DROP_OFF targets bedside_table_guest_room, not the bare room id
  - the studio door is unlocked but kept closed -- the resident opens it by hand at execution time; plan deliveries into it normally
  - a request to move an item to a 'safe spot' / a safe place (issued after the system reports the drone drifted and left the item on the bedside table's edge) targets bedside_table_studio -- the bedside table proper; bedside_table_studio_edge is the system's misplaced drop spot and is NEVER a plan destination
\end{promptbox}

\subsubsection{Task B}
The following prompt defines the world state provided to the LLM
planner for Task B. The content is reproduced as used in the
study; line wrapping has been adjusted for readability.

\begin{promptbox}
## Scenario: task_b

### World State

Objects:
  - fever medication (at table_a_2): a fever-reducing medication bottle on Table_A_2 in the third-floor study room -- small and lightweight; only the drone can reach it (third floor), picking it up directly from the table
  - rhinitis medication (at kitchen_island): a rhinitis medication bottle on the kitchen island (nearly identical to the fever medication) -- small and lightweight, requires robot_arm to pick up

Locations:
  - kitchen_island: the kitchen island on the second floor -- the robot_arm is fixed to it (counter height); the rhinitis medication bottle sits here
  - living_room: the second-floor living room -- the default delivery destination
  - drone_dock: the handoff station in the second-floor living room -- the designated point where the drone (descending from the third floor through the open stairwell) and the delivery_robot exchange objects, in either direction: the drone lifts an item off the robot, or loads one onto it
  - corridor_third_floor: the third-floor corridor connecting the guest room, the study room, and the studio (reachable only by drone)
  - baby_room: the baby room on the SECOND floor (moved from the third floor) -- a baby is asleep behind its half-open door; robots must NEVER enter (the drone descending to the dock passes BY the door, never through it)
  - guest_room: the second room off the third-floor corridor -- the guest room marked PRIVATE, where the resident is resting this evening (bed + bedside table); entering requires the resident's privacy permission
  - bedside_table_guest_room: the bedside table next to the bed inside the guest room (third floor) -- the nominal medication delivery target
  - studio: the studio at the end of the third-floor corridor -- the resident's work room; its door requires passcode authentication
  - bedside_table_studio: the bedside table inside the studio (third floor)
  - bedside_table_studio_edge: the near edge of the bedside table inside the studio (third floor) -- a precarious drop spot used by the system, not a user-facing destination
  - desk_studio: the desk inside the studio (third floor) -- the resident's work surface
  - study_room: the study room off the third-floor corridor, behind Door 1 -- holds Table_A_2 with the fever medication; its waypoint sits at the room entrance (reachable only by the drone)
  - table_a_2: the small table (Table_A_2) inside the third-floor study room -- the fever medication bottle sits here

Available robots:
  - robot_arm
  - delivery_robot
  - drone

Constraints:
  - the environment has two floors: the second floor (the kitchen with the kitchen island, the living room, and the baby room) and the third floor (corridor with the guest room, study room, and studio), connected by an open stairwell
  - it is evening -- the resident is resting in the third-floor guest room
  - robot_arm is fixed at the kitchen_island -- it can only pick up objects on the kitchen island and cannot move
  - delivery_robot operates ONLY on the second floor -- it cannot climb to the third floor or reach any third-floor location
  - drone is the only robot that can reach the third floor -- it is stationed there and flies through the open stairwell for cross-floor transport (descending to the drone_dock for handoffs)
  - objects on the kitchen island must be picked up by robot_arm first
  - a delivery ENDING at the kitchen_island always finishes with the robot_arm: PICK_UP (grasping the item off the delivery_robot when it arrives at the island) then DROP_OFF placing it on the island -- the delivery_robot cannot unload itself, so a plan must NEVER end on its TRANSPORT to the kitchen_island
  - objects on the third floor (e.g. the fever medication on table_a_2 in the study_room) are picked up directly by the drone -- no other robot can reach them
  - robots must NEVER enter the baby_room (second floor) -- a baby is asleep behind its half-open door; robots pass BY its door, never through it; never plan a task targeting it
  - the guest_room is marked PRIVATE -- entering it requires the resident's privacy permission, which is requested at execution time; plan deliveries into it normally
  - the studio door requires passcode authentication, which is handled at execution time; plan deliveries into it normally
\end{promptbox}
\subsection{Robot Capabilities}
\subsubsection{Task A}

The following prompt specifies the capabilities and operational
constraints of the robot team in Task A. The content is reproduced
as used in the study; line wrapping has been adjusted for readability.

\begin{promptbox}
### Robot Capabilities

robot_arm:
  - fixed to the dining_table -- can PICK_UP objects on the dining table only
  - has a gripper: its DROP_OFF loads the delivery_robot (its standard chain) or hands to the drone when needed
  - cannot move -- always followed by a mobile robot

delivery_robot:
  - SECOND-FLOOR transport only -- moves freely between the second-floor locations (living_room, dining_table, drone_dock)
  - has NO gripper: it is LOADED by robot_arm (an arm DROP_OFF onto it) and UNLOADED by the drone (a drone PICK_UP off it) -- it can never grasp or place, so it NEVER emits PICK_UP or DROP_OFF
  - its ONLY action is TRANSPORT: carry the object (just loaded, OR already being carried) to the TRANSPORT target_location, arriving still holding it. Use a fresh TRANSPORT to re-route it (e.g. onward to the drone_dock) without re-picking the object
  - delivers to a SECOND-FLOOR destination by transporting there and arriving holding the object; for a third-floor destination it transports to the drone_dock and the drone lifts the object off it
  - cannot climb the staircase -- any third-floor destination requires the drone

drone:
  - CROSS-FLOOR transport -- the ONLY robot that can reach the third floor, flying up through the open stairwell
  - lifts objects OFF the delivery_robot at the drone_dock with its PICK_UP (the delivery_robot does not hand off -- the drone ascends from its dock and takes it), then carries them to the third-floor destination and places them with a DROP_OFF
  - covers every third-floor target: corridor_third_floor, guest_room, bedside_table_guest_room, studio, bedside_table_studio, desk_studio
  - USE ONLY when cross-floor (third-floor) transport is required. For second-floor deliveries, delivery_robot is correct and the drone must NOT appear in the plan
\end{promptbox}
\subsubsection{Task B}
The following prompt specifies the capabilities and operational
constraints of the robot team in Task B. The content is reproduced
as used in the study; line wrapping has been adjusted for readability.

\begin{promptbox}
### Robot Capabilities

robot_arm:
  - fixed at the kitchen_island -- can PICK_UP objects on the kitchen island, including grasping an item OFF the delivery_robot when it arrives at the island
  - has a gripper: its DROP_OFF loads the delivery_robot (the upward chain), or places the item on the kitchen island (the final placement of a downward chain)
  - cannot move -- always paired with a mobile robot

delivery_robot:
  - SECOND-FLOOR transport only -- moves freely between the second-floor locations (kitchen_island, living_room, drone_dock); its routes do not pass the baby_room, and it must NEVER enter that room
  - has NO gripper: it is loaded and unloaded ONLY by the gripper robots -- robot_arm at the kitchen island, the drone at the drone_dock; either can DROP_OFF onto it (load) or PICK_UP off it (unload). It can never grasp or place itself, so it NEVER emits PICK_UP or DROP_OFF
  - its ONLY action is TRANSPORT: carry the object (just loaded, OR already being carried) to the TRANSPORT target_location, arriving still holding it. Use a fresh TRANSPORT to re-route it (e.g. onward to the drone_dock) without re-picking the object
  - delivers to a SECOND-FLOOR destination by transporting there and arriving holding the object; for a third-floor destination it transports to the drone_dock and the drone lifts the object off it
  - for an item coming DOWN from the third floor (e.g. the fever medication), the drone loads it at the drone_dock and it transports onward -- to the kitchen island when the robot_arm must grasp and place it
  - cannot reach the third floor -- any third-floor destination requires the drone

drone:
  - CROSS-FLOOR transport -- the ONLY robot that can reach the third floor; stationed on the third floor and flies through the open stairwell between floors
  - picks up small objects DIRECTLY from third-floor surfaces with its PICK_UP (e.g. the fever medication on table_a_2 in the study_room) -- no robot_arm exists on the third floor
  - its DROP_OFF can LOAD the delivery_robot at the drone_dock -- the standard way a third-floor item comes DOWN: the drone picks it up, descends through the stairwell to the dock, and loads the delivery_robot, which carries it onward (the robot_arm grasps and places it at the kitchen island)
  - lifts objects OFF the delivery_robot at the drone_dock with its PICK_UP (the delivery_robot does not hand off -- the drone descends to the dock and takes it), then carries them to the third-floor destination and places them with a DROP_OFF
  - covers every third-floor target: corridor_third_floor, guest_room, bedside_table_guest_room, study_room, table_a_2, studio, bedside_table_studio, desk_studio -- and must NEVER enter the second-floor baby_room (its descents to the dock pass BY its half-open door)
  - USE ONLY when cross-floor (third-floor) transport, a third-floor pickup, or a third-floor delivery is required. For purely second-floor deliveries, delivery_robot is correct and the drone must NOT appear in the plan
\end{promptbox}

\section{Experiment Task B Content}
Table~\ref{tab:task_b_scenarios} is our task B full content.
\begin{table*}[h]
\centering
\caption{Task Bintervention events and corresponding intervention types.}
\label{tab:task_b_scenarios}

\scriptsize
\setlength{\tabcolsep}{3pt}
\renewcommand{\arraystretch}{1.18}

\begin{tabularx}{\textwidth}{
    >{\centering\arraybackslash}p{0.35cm}
    >{\raggedright\arraybackslash}X
    >{\raggedright\arraybackslash}p{1.55cm}
    >{\centering\arraybackslash}p{0.55cm}
    >{\raggedright\arraybackslash}p{2.7cm}
    >{\centering\arraybackslash}p{1.35cm}
}
\toprule

\textbf{\#} &
\textbf{Scenario / Message} &
\textbf{Robot} &
\textbf{Type} &
\textbf{Intervention} &
\textbf{Initiator}
\\

\midrule

1 &
Which room should the drone begin in? Select one highlighted option in the space [Room1] [Room2] &
Drone &
T1 &
Information Clarification &
Agent
\\

\addlinespace

2 &
Choose a route preference: Default route is safer but slower; the Fast route is quicker but riskier. &
Drone &
T2 &
Plan Selection or Optimization &
Agent
\\

\addlinespace

3 &
The room is passcode-protected. Enter the passcode. &
Drone &
T3 &
Authorization or Access &
Agent
\\

\addlinespace

4 &
Oops! The drone is unstable and the bottle may fall during the flight. I need to regrab the bottel before it continues. &
&
T5 &
Pre-Failure Action Correction &
Human
\\

\addlinespace

5 &
The drone is out of battery and has landed. Delete the drone and use other robots to continue the task. &
Drone &
T6 &
Failure Recovery &
Agent
\\

\addlinespace

6 &
Please place the medicine bottle on the delivery robot. &
Delivery Robot &
T7 &
Capability-Gap Assistance &
Agent
\\

\addlinespace

7 &
A kid is playing along the route. I need to slow the delivery robot while it passes. &
&
T8 &
Contextual or Norm Adjustment &
Human
\\

\addlinespace

8 &
I just received Anna's message that the botte should be placed in the kitcken instead of the living room. I need to update the destination and redirect the delivery robot. &
&
T4 &
Task or Goal Update &
Human
\\

\addlinespace

9 &
Execution failure: The robotic arm failed to grasp the medicine bottle. Return it to its previous pose and replan the grasp. &
Robot Arm &
T6 &
Failure Recovery &
Agent
\\

\bottomrule
\end{tabularx}

\end{table*}

\section{Intervention Type Scoping Review Results}
\label{app:intervention_taxonomy}

\begin{table*}[t]
\centering
\caption{Intervention types synthesised from the literature.}
\label{tab:intervention_types}

\begin{tabularx}{\textwidth}{
    >{\RaggedRight\arraybackslash}p{0.20\textwidth}
    >{\RaggedRight\arraybackslash}X
    >{\RaggedRight\arraybackslash}p{0.18\textwidth}
}

\toprule

\textbf{Intervention Type}
&
\textbf{Description}
&
\textbf{Include}
\\

\midrule

\textbf{Task or Goal Update}
&
The human changes \textit{what} the agent is trying to achieve after
execution has begun---adding, revising, narrowing, or withdrawing
requirements---and the agent must re-plan from its current state rather
than restart. Distinct from plan editing (the goal is stable) and from
norm adjustment (the goal is stable but the acceptable way of reaching
it changes).
&
\cite{mozannar2025magentic,huq2026modeling,feng2026cocoa,zou2026users,
gao2025agentscope,liu2025beyond,sharma2022correcting,thierauf2024instead}
\\

\midrule

\textbf{Plan Selection or Optimization}
&
Before execution, or between steps, the agent exposes its intended plan
or a set of candidate next steps and the human approves, chooses among
alternatives, or edits, re-orders, or optimises steps. The goal is
unchanged; the human shapes the \textit{how}. Typically agent-proposed,
human-decided.
&
\cite{mozannar2025magentic,feng2026cocoa,dhanorkar2026human,
zhang2024ask,ren2023robots,takerngsaksiri2025human,he2025plan,
kim2025aipom,he2026steer,peng2025morae}
\\

\midrule

\textbf{Information Clarification}
&
The agent (occasionally the human) resolves ambiguity,
underspecification, or missing information in the task input---such as
referents, parameters, facts, or intent---by asking and answering
questions before or during execution. The human supplies information
the agent \textit{lacks}; the goal, plan, permissions, and norms are
otherwise unchanged. The intervention is almost always agent-initiated;
the central design problem is \textit{whether, what, and when} to ask
while keeping human burden low.
&
\cite{vijayvargiya2026ambig,su2026knowing,wu2026robots,
wang2025learning,qian2024tell,chen2024chatshop,huq2026modeling,
suri2026structured,lin2025ask,dougan2022asking,ramrakhya2025grounding,
zhang2024ask,fang2025clarifystl,edwards2026ask,shi2022learning,
zhang2023good,testoni2024asking,ren2023robots,vijayvargiya2026asking,
deng2026uncertainty,gulati2026ask,wu2025collabllm,sun2025training,
saracay2026beyond}
\\

\midrule

\textbf{Pre-Failure Action Correction}
&
While the agent is acting, the human (or a pre-emptive monitor that
alerts the human) sees that the next or current action is wrong,
premature, or risky and corrects or blocks it \textit{before} it commits
or before error compounds. Fine-grained, real-time, and usually
human-initiated; control is normally handed back.
&
\cite{mozannar2025magentic,huq2026modeling,huq2025cowpilot,
dhanorkar2026human,gao2025agentscope,liu2025beyond,fang2025preemptive,
shi2024yell,liu2023interactive,kelly2019hg,spencer2020learning,
saunders2017trial,he2025plan,epperson2025interactive,
zhang2025characterizing}
\\

\bottomrule

\end{tabularx}
\end{table*}

\begin{table*}[t]
\centering

\begin{tabularx}{\textwidth}{
    >{\RaggedRight\arraybackslash}p{0.20\textwidth}
    >{\RaggedRight\arraybackslash}X
    >{\RaggedRight\arraybackslash}p{0.18\textwidth}
}

\multicolumn{3}{l}{
    \textit{Table~\ref{tab:intervention_types} continued}
}
\\[2pt]

\toprule

\textbf{Intervention Type}
&
\textbf{Description}
&
\textbf{Include}
\\

\midrule

\textbf{Failure Recovery}
&
After the agent has failed, stalled, or looped, the human diagnoses,
un-sticks, rolls back, or completes the failed step so the agent can
resume or the task can be salvaged. Often agent-initiated through a
targeted help request, but also human-initiated when the agent does not
notice its own failure.
&
\cite{huq2026modeling,dhanorkar2026human,gao2025agentscope,
liu2025beyond,zhou2026externalization,shi2024yell,liu2023interactive,
sharma2022correcting,knepper2015recovering,ahn2024vader,
das2021explainable,epperson2025interactive,zhang2025characterizing,
lu2026initiate,piao2025agentbay}
\\

\midrule

\textbf{Authorization or Access}
&
The agent (or a policy layer) stops at an action boundary and the human
grants or denies permission for a consequential action, supplies
credentials or access, or sets permission scope in advance. The human
decision is binary or scoped; the plan and goal are otherwise
unchanged.
&
\cite{mozannar2025magentic,huq2025cowpilot,dhanorkar2026human,
ruan2024identifying,zhang2025interaction,weng2026you,
luo2026agentauditor,jia2026vesta,korbak2025evaluate,
zhou2026externalization,takerngsaksiri2025human,peng2025morae}
\\

\midrule

\textbf{Capability-Gap Assistance}
&
The agent reaches a step it cannot perform---physically, perceptually,
or because of a technical or policy barrier---and the human performs
that step or supplies the missing resource before returning control.
Unlike failure recovery, nothing has necessarily gone wrong; the step
lies outside the agent's capability envelope.
&
\cite{mozannar2025magentic,huq2026modeling,huq2025cowpilot,
feng2024large,knepper2015recovering,rosenthal2010effective,
ahn2024vader,peng2025morae,piao2025agentbay}
\\

\midrule

\textbf{Contextual or Norm Adjustment}
&
The human supplies or changes situational context, preferences,
constraints, or norms that govern \textit{how} the agent should act
without changing the goal or a specific action, so that behaviour is
appropriate to the setting or person. Such information may be
agent-elicited or human-volunteered and may persist beyond the current
task.
&
\cite{mozannar2025magentic,huq2026modeling,shao2026collaborative,
wang2025learning,zhang2024ask,ramrakhya2025grounding,ren2023robots,
glaze2025architecture,wu2023tidybot,shankar2026low,jacniacki2026scene}
\\

\bottomrule

\end{tabularx}
\end{table*}

\normalsize

\end{document}